\documentclass[twocolumn,english,prb,citeautoscript,superscriptaddress]{revtex4-1}
\usepackage[T1]{fontenc}
\usepackage[latin9]{inputenc}
\usepackage{color}
\usepackage{babel}
\usepackage{amsmath}
\usepackage{amssymb}
\usepackage{graphicx}
\usepackage{esint}
\usepackage[unicode=true,pdfusetitle,
 bookmarks=true,bookmarksnumbered=false,bookmarksopen=false,
 breaklinks=false,pdfborder={0 0 1},backref=false,colorlinks=true]
 {hyperref}

\makeatletter

\usepackage{babel}

\usepackage{amsfonts}\usepackage{bbm}\usepackage{graphics}

\def\ket#1{\left| #1\right>}
\def\bra#1{\left< #1\right|}

\def\<{\langle}
\def\>{\rangle}

\usepackage{leftidx}

\usepackage{babel}

\usepackage{hyperref}
\hypersetup{colorlinks=true,citecolor=blue}

\makeatother

\begin{document}

\title{Capacitively coupled singlet-triplet qubits in the double charge
resonant regime}

\author{V. Srinivasa}

\email{vsriniv@umd.edu}

\affiliation{Joint Quantum Institute, University of Maryland, College Park, Maryland
20742, USA}

\affiliation{National Institute of Standards and Technology, Gaithersburg, Maryland
20899, USA}

\affiliation{Laboratory for Physical Sciences, College Park, Maryland 20740, USA}

\author{J. M. Taylor}

\affiliation{Joint Quantum Institute, University of Maryland, College Park, Maryland
20742, USA}

\affiliation{National Institute of Standards and Technology, Gaithersburg, Maryland
20899, USA}

\affiliation{Joint Center for Quantum Information and Computer Science, University
of Maryland, College Park, Maryland 20742, USA}
\begin{abstract}
We investigate a method for entangling two singlet-triplet qubits
in adjacent double quantum dots via capacitive interactions. In contrast
to prior work, here we focus on a regime with strong interactions
between the qubits. The interplay of the interaction energy and simultaneous
large detunings for both double dots gives rise to the ``double charge
resonant'' regime, in which the unpolarized (1111) and fully polarized
(0202) four-electron states in the absence of interqubit tunneling
are near degeneracy, while being energetically well-separated from
the partially polarized (0211 and 1102) states. A rapid controlled-phase
gate may be realized by combining time evolution in this regime in
the presence of intraqubit tunneling and the interqubit Coulomb interaction
with refocusing $\pi$ pulses that swap the singly occupied singlet
and triplet states of the two qubits via, e.g., magnetic gradients.
We calculate the fidelity of this entangling gate, incorporating models
for two types of noise -- charge fluctuations in the single-qubit
detunings and charge relaxation within the low-energy subspace via
electron-phonon interaction -- and identify parameter regimes that
optimize the fidelity. The rates of phonon-induced decay for pairs
of GaAs or Si double quantum dots vary with the sizes of the dipolar
and quadrupolar contributions and are several orders of magnitude
smaller for Si, leading to high theoretical gate fidelities for coupled
singlet-triplet qubits in Si dots. We also consider the dependence
of the capacitive coupling on the relative orientation of the double
dots and find that a linear geometry provides the fastest potential
gate. 
\end{abstract}
\maketitle

\section{Introduction}

Electrons spins confined within semiconductor quantum dots form the
basis of a highly controllable and potentially scalable approach to
solid-state quantum information processing \cite{Loss1998,Taylor2005,Hanson2007RMP,Kloeffel2013,Zwanenburg2013}.
The encoding of spin quantum bits (qubits) in two-electron singlet
and triplet states of a double quantum dot \cite{Burkard1999,Levy2002,Taylor2005,Taylor2007}
enables rapid, universal manipulation via tuning of the singlet-triplet
(exchange) splitting through electrical control over the double-dot
potential \cite{Petta2005} combined with static magnetic field gradients
\cite{Pioro-Ladriere2008,Foletti2009}, without requiring time-dependent
magnetic fields and while simultaneously providing protection against
errors induced by hyperfine interaction \cite{Johnson2005,Taylor2005PRL,Koppens2005,Coish2005,Petta2005,Laird2006,Taylor2007,Bluhm2011,Wang2012}.
Coherent control of singlet-triplet qubits has been experimentally
demonstrated in the context of both single-qubit manipulation \cite{Petta2005,Foletti2009,Barthel2010,Bluhm2011,Maune2012,Wu2014}
and two-qubit entanglement \cite{vanWeperen2011,Shulman2012}. 

For a pair of singlet-triplet qubits coupled via tunneling, the effective
exchange interaction can be used to carry out two-qubit gates \cite{Levy2002,Klinovaja2012,Li2012,Kestner2013,Wardrop2014,Mehl2014};
however, this approach typically requires an accompanying mechanism
for suppressing errors due to leakage out of the qubit subspace during
gate operation. Alternatively, two singlet-triplet qubits in adjacent
double dots may be entangled via capacitive coupling \cite{vanderWiel2002,Taylor2005,Hanson2007PRL,Stepanenko2007,Ramon2011,vanWeperen2011,Trifunovic2012,Shulman2012,Nielsen2012,Wang2014arxiv,Calderon-Vargas2015}.
In this case, interqubit tunneling is absent and the entanglement
instead originates from the Coulomb interaction of the multipole moments
associated with the different charge distributions of the singlet
and triplet states \cite{Coish2005}. The spin-dependent charge dipole
moments of spatially separated singlet-triplet qubits can also be
coupled to microwaves, enabling long-range, high-frequency gating
\cite{Burkard2006,Taylor2006}. Nevertheless, realizing robust entangling
gates in the presence of the charge-based decoherence mechanisms typically
existing in the solid state, including both dephasing \cite{Barrett2002,Coish2005,Hu2006,Taylor2007,Culcer2009,Dial2013}
and relaxation via, e.g., coupling to phonons \cite{Brandes2002,Johnson2005,Taylor2007,Meunier2007,Barthel2012,Raith2012PRB,Danon2013,Braakman2014},
remains challenging. 

Here, we consider a pair of capacitively coupled singlet-triplet qubits
in the absence of interqubit tunneling. In contrast to the repulsive
interqubit dipole-dipole interaction originally considered in Ref.
\citenum{Taylor2005}, we focus specifically on the case of an attractive
dipole-dipole interaction, implemented by adjusting via external gate
voltages the energy detunings between the singly and doubly occupied
two-electron charge configurations such that they are large for both
double dots. The interplay of these large detunings and the Coulomb
interaction energy gives rise to the ``double charge resonant''
regime, as we describe below.

Combining time evolution in this regime with single-qubit $\pi$ pulses
that swap the singly occupied singlet and triplet states of both qubits
using, e.g., static magnetic gradients \cite{Shulman2012} leads to
a controlled $\pi$-phase (or controlled-Z) entangling gate. As a
consequence of the attractive dipole-dipole interaction, increasing
the speed of this gate simultaneously decreases the gate error due
to charge noise. We calculate the gate fidelity in the presence of
charge fluctuations in the double-dot detunings and identify gate
voltages and coupling strengths at which the fidelity is optimized.
We then investigate charge relaxation due to electron-phonon coupling
for both GaAs and Si double quantum dots in linear and purely quadrupolar
dot configurations and determine the effects of both this relaxation
and fast charge noise on the gate fidelity. Finally, we consider the
geometry dependence of the interqubit capacitive coupling and identify
the linear geometry as a configuration that maximizes the gate speed.

\section{Model and double charge resonant regime\label{sec:Model}}

\begin{figure}
\includegraphics[bb=-10bp 0bp 360bp 330bp,width=3.3in]{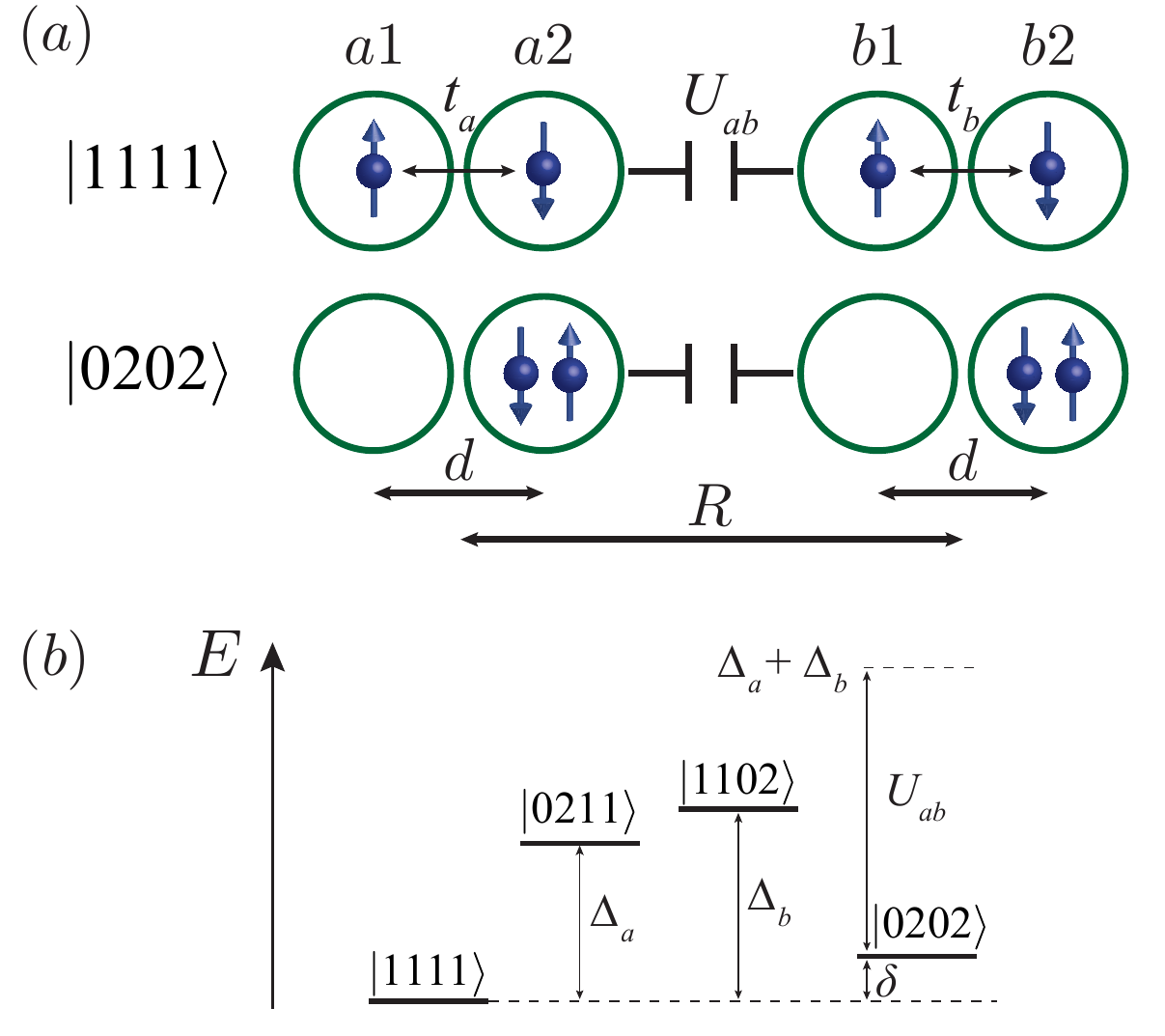}

\protect\caption{\label{fig:doubledouble}(a) Schematic diagram of capacitively coupled
double quantum dots in the charge states $\ket{1111}$ and $\ket{0202}.$
The interdot spacing within each double dot is $d,$ and the separation
between the centers of the double dots is $R.$ (b) Energy level diagram
for the main four-electron charge configurations considered in the
present work, illustrating the double charge resonant regime. }
\end{figure}

We consider two singlet-triplet qubits, realized within a pair of
adjacent two-electron double quantum dots {[}Fig. \ref{fig:doubledouble}(a){]}
with only the lowest orbital level of each dot taken into account.
Each two-electron double dot encodes one qubit. As in Ref. \citenum{Taylor2005},
we initially assume a linear geometry in which the tunnel barriers
are adjusted via gates such that tunneling occurs only between the
dots within each qubit, while adjacent dots belonging to different
qubits are coupled purely capacitively. We can write a Hubbard Hamiltonian
for the system \cite{Yang2011Hubbard} as $H_{{\rm hub}}=H_{a}+H_{b}+H_{{\rm int}},$
where

\begin{eqnarray}
H_{\alpha} & = & H_{\alpha n}+H_{\alpha t},\label{eq:Halpha}\\
H_{\alpha n} & = & \sum_{i=1,2}\left[\epsilon_{\alpha i}n_{\alpha i}+\frac{U_{\alpha}}{2}n_{\alpha i}(n_{\alpha i}-1)\right]\nonumber \\
 &  & +\ V_{\alpha}n_{\alpha1}n_{\alpha2},\label{eq:Halphan}\\
H_{\alpha t} & = & \sum_{i\neq j}\sum_{\sigma}t_{\alpha}c_{\alpha i\sigma}^{\dag}c_{\alpha j\sigma},\label{eq:Halphat}
\end{eqnarray}
is the Hamiltonian for double dot $\alpha=a,b,$ and $H_{{\rm int}}$
is the capacitive interaction between the double dots. For simplicity,
we initially include only the dominant interaction term for the linear
geometry we consider, 
\begin{eqnarray}
H_{{\rm int}} & = & U_{ab}n_{a2}n_{b1}.\label{eq:Hint}
\end{eqnarray}
Equations \eqref{eq:Halphan} and \eqref{eq:Hint} are expressed in
terms of the electron number operators $n_{\alpha i}=\sum_{\sigma}n_{\alpha i\sigma}=\sum_{\sigma}c_{\alpha i\sigma}^{\dagger}c_{\alpha i\sigma},$
where $c_{\alpha i\sigma}^{\dagger}$ creates an electron in dot $i$
of qubit $\alpha$ with spin $\sigma=\uparrow,\downarrow$ and orbital
energy $\epsilon_{\alpha i}.$ These terms determine the energy of
each four-electron charge configuration $\ket{n_{a1}\ n_{a2}\ n_{b1}\ n_{b2}}$
in the absence of interdot tunneling. The quantities $U_{\alpha}$
and $V_{\alpha}$ are the Coulomb repulsion energies for two electrons
in the same dot and in different dots within qubit $\alpha,$ respectively.
$H_{\alpha t}$ couples the double-dot charge configurations $\left(n_{\alpha1},n_{\alpha2}\right)$
via tunneling, and $t_{\alpha}$ denotes the tunneling amplitude for
double dot $\alpha.$ As discussed in Refs. \citenum{Taylor2005}
and \citenum{Taylor2007}, each double dot can be described in the
two-electron regime as an effective three-level system with a state
space spanned by 
\begin{eqnarray}
\ket{T_{11}} & \equiv & \ket{\left(1,1\right)T_{0}}=\frac{1}{\sqrt{2}}\left(c_{1\uparrow}^{\dagger}c_{2\downarrow}^{\dagger}+c_{1\downarrow}^{\dagger}c_{2\uparrow}^{\dagger}\right)\ket{0},\label{eq:T11}\\
\ket{S_{11}} & \equiv & \ket{\left(1,1\right)S}=\frac{1}{\sqrt{2}}\left(c_{1\uparrow}^{\dagger}c_{2\downarrow}^{\dagger}-c_{1\downarrow}^{\dagger}c_{2\uparrow}^{\dagger}\right)\ket{0},\label{eq:S11}\\
\ket{S_{02}} & \equiv & \ket{\left(0,2\right)S}=c_{2\uparrow}^{\dagger}c_{2\downarrow}^{\dagger}\ket{0},\label{eq:S02}
\end{eqnarray}
where the qubit index $\alpha$ has been suppressed for clarity. 

In our analysis of the capacitively coupled double-dot pair system,
we focus on the four-electron charge subspaces $\ket{1111},$ $\ket{0202},$
$\ket{1102},$ and $\ket{0211}.$ Noting that $H_{{\rm hub}}$ conserves
both the total spin and the total $z$ component of spin and that
$H_{\alpha t}$ couples only the two-electron singlet states $\ket{S_{11}}$
and $\ket{S_{02}}$ within double dot $\alpha,$ we may consider the
subspace spanned by product states of the form $\ket{S_{a},S_{b}}\equiv\ket{S_{a}}\otimes\ket{S_{b}},$
where $\ket{S_{\alpha}}\in\left\{ \ket{S_{11}},\ket{S_{02}}\right\} $
for $\alpha=a,b.$ In the basis $\left\{ \ket{S_{11},S_{11}},\ket{S_{02},S_{02}},\ket{S_{11},S_{02}},\ket{S_{02},S_{11}}\right\} ,$
the Hamiltonian has the representation 
\begin{equation}
H_{{\rm hub}}=\left(\begin{array}{cccc}
0 & 0 & \sqrt{2}t_{b} & \sqrt{2}t_{a}\\
0 & \delta & \sqrt{2}t_{a} & \sqrt{2}t_{b}\\
\sqrt{2}t_{b} & \sqrt{2}t_{a} & \Delta_{b} & 0\\
\sqrt{2}t_{a} & \sqrt{2}t_{b} & 0 & \Delta_{a}
\end{array}\right),\label{eq:Hhubmat}
\end{equation}
where $\Delta_{a}\equiv-\epsilon_{a}+U_{a}-V_{a}+U_{ab}$ and $\Delta_{b}\equiv-\epsilon_{b}+U_{b}-V_{b}-U_{ab}$
are the effective energy detunings between $\ket{S_{11}}$ and $\ket{S_{02}}$
for double dots $a$ and $b,$ respectively (accounting for coupling
to the other double dot), $\epsilon_{\alpha}\equiv\epsilon_{\alpha1}-\epsilon_{\alpha2},$
and $\delta\equiv\Delta_{a}+\Delta_{b}-U_{ab}$ is the energy difference
between $\ket{S_{11},S_{11}}$ and $\ket{S_{02},S_{02}}.$ The detunings
$\Delta_{\alpha}$ are controlled via tuning of the on-site energies
via gate voltages, which set $\epsilon_{a}$ and $\epsilon_{b}.$ 

The controlled-phase gate for two singlet-triplet qubits discussed
in Sec. \ref{sec:CPHASEgate} involves tunneling from $\ket{S_{11}}$
to $\ket{S_{02}}$ for $\alpha=a,b$, which simultaneously induces
dipole moments in both double dots. Given that the interqubit Coulomb
interaction strength $U_{ab}>0,$ the regime of interest for the operation
of this gate is that in which $\Delta_{\alpha}\gg\delta>0$ for $\alpha=a,b,$
so that $\ket{S_{11},S_{11}}$ is lower in energy than $\ket{S_{02},S_{02}}$,
while $\ket{S_{11},S_{11}}$ and $\ket{S_{02},S_{02}}$ are energetically
well-separated from $\ket{S_{11},S_{02}}$ and $\ket{S_{02},S_{11}}$
{[}Fig. \ref{fig:doubledouble}(b){]}. We refer to this regime as
the ``double charge resonant regime,'' as the attractive interaction
between the dipole moments of the two double dots in the state $\ket{S_{02},S_{02}}$
effectively brings it into near-resonance with $\ket{S_{11},S_{11}}.$
Note that this regime is not accessible in the scenario originally
studied in Ref. \citenum{Taylor2005}, where the state $\ket{S_{02},S_{20}}$
is considered instead of $\ket{S_{02},S_{02}}$ and the interqubit
Coulomb interaction between the dipole moments is repulsive.

We show in Sec. \ref{sec:CPHASEgate} that, in contrast to the nonresonant
regime of Ref. \citenum{Taylor2005}, the double charge resonant regime
enables a controlled-phase gate to be generated by dynamics within
an effective low-energy subspace derived from $\ket{S_{11},S_{11}}$
and $\ket{S_{02},S_{02}}.$ In order to compare two-qubit phase gates
in the nonresonant and double charge resonant regimes, we now estimate
the scaling of the phase gate errors in the presence of detuning noise.
Using Eq. \eqref{eq:Hhubmat}, we calculate the fourth-order energy
shift for the state $\ket{S_{11},S_{11}}$ due to the interqubit capacitive
coupling (i.e., the additional energy shift for $U_{ab}\neq0$), which
gives the rate of the phase gate. For the nonresonant regime, the
Hamiltonian has the same form as Eq. \eqref{eq:Hhubmat} with the
replacements $\ket{S_{02},S_{02}}\rightarrow\ket{S_{02},S_{20}},$
$\ket{S_{11},S_{02}}\rightarrow\ket{S_{11},S_{20}},$ and $\delta=\Delta_{a}+\Delta_{b}+U_{ab}$
(here, $\Delta_{b}\equiv\epsilon_{b}+U_{b}-V_{b}+U_{ab}$). Assuming
$\Delta_{a}=\Delta_{b}\equiv\Delta$ and $t_{a}=t_{b}\equiv t$ for
simplicity, we find a phase gate rate $\varepsilon_{1111}=8t^{4}\left(\delta-2\Delta\right)/\Delta^{3}\delta.$ 

Setting $\Delta^{\prime}=\Delta+\xi,$ where $\xi$ represents classical,
static, Gaussian-distributed noise in the single-qubit detunings due
to gate voltage fluctuations \cite{Dial2013}, we can write the nontrivial
phase factor acquired by the state derived from $\ket{S_{11},S_{11}}$
for $U_{ab}\neq0$ as $e^{i\phi^{\prime}},$ with $\phi^{\prime}=\phi_{0}+\phi_{\xi}.$
Here, $\phi_{0}\equiv\varepsilon_{1111}\tau_{{\rm gate}}$ is the
phase acquired during the gate time $\tau_{{\rm gate}}$ in the absence
of noise and $\phi_{\xi}\approx\left(4t^{2}/\Delta^{2}\right)\xi\tau_{{\rm gate}}$
represents the phase fluctuations, approximated using the second-order
energy shift. Averaging over the noise gives $\left\langle e^{i\phi^{\prime}}\right\rangle =e^{-\Gamma_{{\rm \xi}}^{2}\tau^{2}}e^{i\phi_{0}}\approx\left(1-\Gamma_{{\rm \xi}}^{2}\tau^{2}\right)e^{i\phi_{0}}$
with $\Gamma_{{\rm \xi}}\sim t^{2}/\Delta^{2}T_{2,{\rm \xi}}^{\ast}$
($T_{2,\xi}^{\ast}$ denotes the dephasing time associated with the
charge fluctuations $\xi$), so that the phase gate error can be approximated
as ${\rm err}\approx\Gamma_{{\rm \xi}}^{2}\tau_{{\rm gate}}^{2}\sim\delta^{2}\Delta^{2}/t^{4}\left(\delta-2\Delta\right)^{2}T_{2,{\rm \xi}}^{\ast2}.$
For the nonresonant regime, $\delta=2\Delta+U_{ab}$ and ${\rm err}\sim\left(\Delta^{2}/t^{2}\right)\left(1+2\Delta/U_{ab}\right)^{2}/\left(tT_{2,{\rm \xi}}^{\ast}\right)^{2},$
while for the double charge resonant regime, $\delta=2\Delta-U_{ab}$
and ${\rm err}\sim\delta^{2}/t^{2}\left(tT_{2,{\rm \xi}}^{\ast}\right)^{2}.$
Thus, the phase gate error due to gate voltage fluctuations has the
scaling $\sim\Delta^{2}/t^{2}$ for the nonresonant regime, which
is unfavorable for suppressing errors arising from dipole transitions
to $\ket{S_{11},S_{20}}$ and $\ket{S_{02},S_{11}}$ by keeping $\Delta/t$
large. On the other hand, the error scales as $\sim\delta^{2}/t^{2}$
in the double charge resonant regime, so that faster gates (corresponding
to stronger coupling $U_{ab}$ and therefore smaller $\delta$ for
fixed $\Delta$) are also associated with smaller error. 

We now proceed with a more detailed analysis of the double charge
resonant regime and consider the low-energy effective Hamiltonian
in the subspace $\left\{ \ket{S_{11},S_{11}},\ket{S_{02},S_{02}}\right\} .$
Applying a Schrieffer-Wolff transformation of the form $\tilde{H}=e^{\lambda A}H_{{\rm hub}}e^{-\lambda A}$
with $\lambda\propto t_{\alpha}$ (assuming $t_{a}\sim t_{b})$ to
the Hamiltonian in Eq. \eqref{eq:Hhubmat}, we choose $A$ such that
the coupling to the higher-energy states $\ket{S_{11},S_{02}}$ and
$\ket{S_{02},S_{11}}$ is eliminated up to $\mathcal{O}\left(\lambda^{2}\right).$
Expressions for the basis states resulting from this transformation
are given in Appendix \ref{sec:SW-basis-states}.

\begin{figure}
\includegraphics[bb=0bp 100bp 550bp 430bp,width=3.3in]{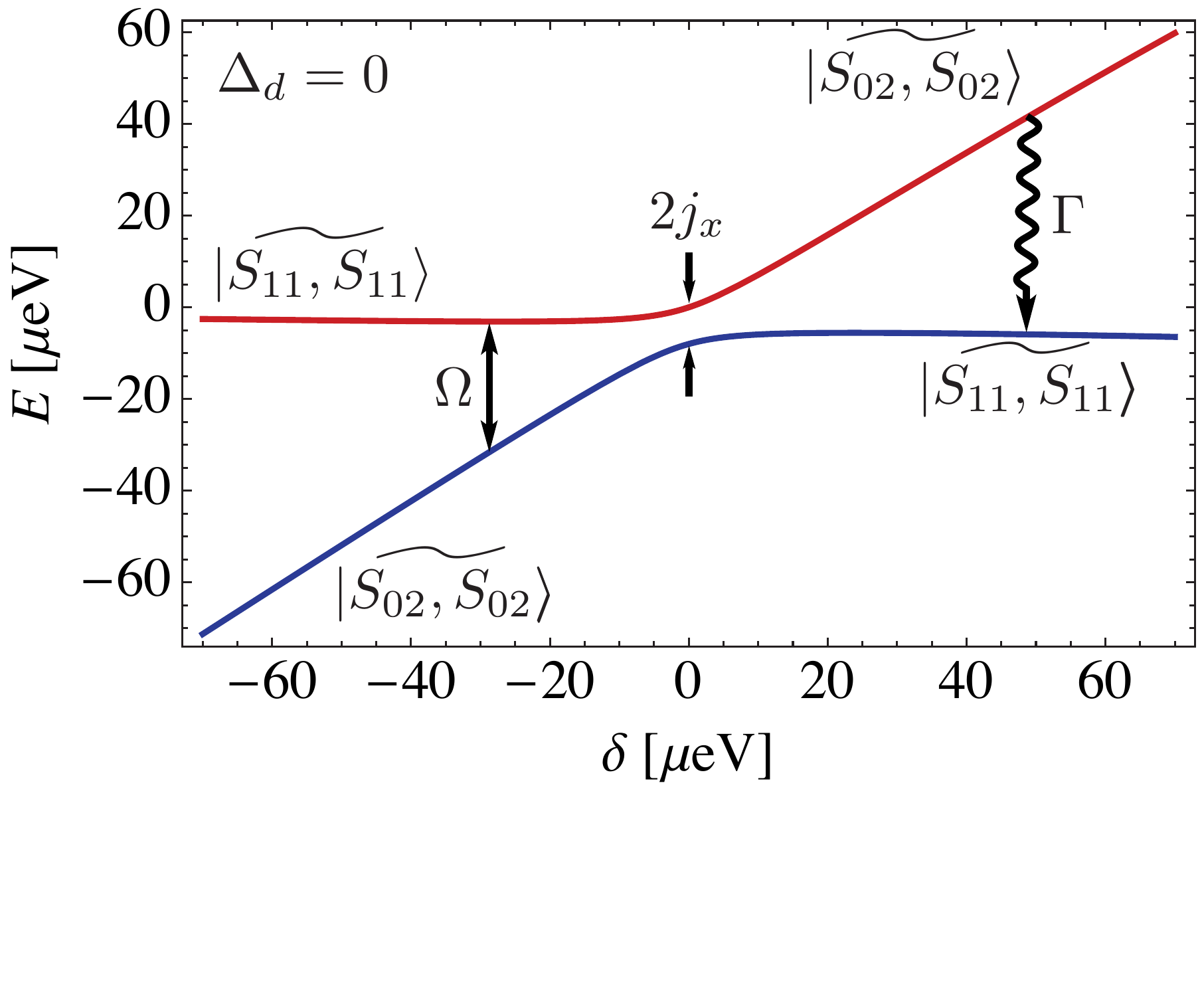}

\protect\caption{\label{fig:energies}Spectrum of $H_{{\rm eff}}$ {[}Eq. \eqref{eq:Heff}{]}
as a function of $\delta$ for $U_{ab}=200\ \mu{\rm eV},$ $\Delta_{d}=0,$
and $\eta_{a}=\eta_{b}\equiv\eta_{0}=0.1.$ }
\end{figure}

Defining $\sigma_{z}=\ket{\widetilde{S_{11},S_{11}}}\bra{\widetilde{S_{11},S_{11}}}-\ket{\widetilde{S_{02},S_{02}}}\bra{\widetilde{S_{02},S_{02}}},$
the effective Hamiltonian within the transformed subspace is 
\begin{equation}
H_{{\rm eff}}=-\left(J_{a}+J_{b}-\frac{j_{d}}{2}\right){\bf 1}-\frac{j_{d}}{2}\sigma_{z}-j_{x}\sigma_{x},\label{eq:Heff}
\end{equation}
where, in terms of the charge admixture parameters $\eta_{\alpha}\equiv t_{\alpha}/\Delta_{\alpha},$
$\delta,$ and the difference of the detunings $\Delta_{d}\equiv\Delta_{a}-\Delta_{b},$

\begin{eqnarray}
J_{a} & \equiv & \eta_{a}^{2}\left(U_{ab}+\delta+\Delta_{d}\right),\label{eq:Ja}
\end{eqnarray}
\begin{eqnarray}
J_{b} & \equiv & \eta_{b}^{2}\left(U_{ab}+\delta-\Delta_{d}\right),\label{eq:Jb}
\end{eqnarray}

\begin{eqnarray}
j_{d} & \equiv & \delta-2\left(J_{a}\frac{\delta+\Delta_{d}}{U_{ab}-\delta-\Delta_{d}}\right.\nonumber \\
 &  & \left.\ \ \ \ \ \ \ +J_{b}\frac{\delta-\Delta_{d}}{U_{ab}-\delta+\Delta_{d}}\right),\label{eq:jd}
\end{eqnarray}

\begin{eqnarray}
j_{x} & \equiv & 2\eta_{a}\eta_{b}U_{ab}\left[1+\delta\left(\frac{1}{U_{ab}-\delta-\Delta_{d}}\right.\right.\nonumber \\
 &  & \left.\left.\ \ \ \ \ \ \ \ \ \ \ \ \ \ \ \ \ \ \ \ \ +\frac{1}{U_{ab}-\delta+\Delta_{d}}\right)\right].\label{eq:jx}
\end{eqnarray}
Diagonalization of Eq. \eqref{eq:Heff} yields the eigenstates $\ket{g}=\cos\theta\ket{\widetilde{S_{11},S_{11}}}-\sin\theta\ket{\widetilde{S_{02},S_{02}}}$
and $\ket{e}=\sin\theta\ket{\widetilde{S_{11},S_{11}}}+\cos\theta\ket{\widetilde{S_{02},S_{02}}},$
where
\begin{eqnarray}
\tan\theta & = & \frac{j_{d}-\Omega}{2j_{x}}
\end{eqnarray}
and $\Omega\equiv E_{e}-E_{g}=\sqrt{j_{d}^{2}+4j_{x}^{2}}$ is the
energy gap between $\ket{g}$ and $\ket{e}.$ The spectrum of $H_{{\rm eff}}$
is shown in Fig. \ref{fig:energies} as a function of $\delta$ for
$U_{ab}=200\ \mu{\rm eV},$ $\Delta_{d}=0,$ and $\eta_{a}=\eta_{b}\equiv\eta_{0}=0.1.$
An avoided crossing occurs at $\delta=0,$ i.e., when $\ket{S_{11},S_{11}}$
and $\ket{S_{02},S_{02}}$ are resonant. For $\delta\gg0,$ $\ket{g}\approx\ket{\widetilde{S_{11},S_{11}}}$
and $\ket{e}\approx\ket{\widetilde{S_{02},S_{02}}}.$

\section{Controlled-phase gate\label{sec:CPHASEgate}}

The time evolution generated by the Hamiltonian $H_{{\rm eff}}$ {[}Eq.
\eqref{eq:Heff}{]} within the two-singlet subspace spanned by $\left\{ \ket{\widetilde{S_{11},S_{11}}},\ket{\widetilde{S_{02},S_{02}}}\right\} $
leads to a controlled-phase gate between the two singlet-triplet qubits
that is based on capacitive coupling in the double charge resonant
regime. In order to obtain this two-qubit entangling gate, we now
incorporate the triplet states of the double dots {[}Eq. \eqref{eq:T11}{]}
into the analysis. Since the full sequence for the controlled-phase
gate also involves single-qubit rotations around two orthogonal axes,
we consider the Hamiltonian $H_{{\rm hub}}+H_{aZ}+H_{bZ},$ where
\begin{equation}
H_{\alpha Z}=\frac{g\mu_{B}}{2}\sum_{i=1,2}\sum_{\sigma,\sigma^{\prime}}c_{\alpha i\sigma}^{\dagger}\left({\bf B}_{\alpha i}\cdot\boldsymbol{\sigma}_{\alpha}\right)c_{\alpha i\sigma^{\prime}}\label{eq:HalphaZ}
\end{equation}
represents Zeeman coupling to magnetic fields ${\bf B}_{\alpha i}$
for double dot $\alpha$ (here, $g$ is the effective electron $g$
factor and $\mu_{B}$ denotes the Bohr magneton). Combining this Zeeman
coupling with the spin-independent Hubbard term $H_{\alpha}$ {[}Eqs.
\eqref{eq:Halpha}-\eqref{eq:Halphat}{]} that leads to the exchange
$J_{\alpha}$ enables universal one-qubit control of singlet-triplet
qubits \cite{Levy2002,Petta2005,Taylor2005,Taylor2007,Foletti2009,Barthel2010,Bluhm2011,Maune2012,Wu2014}.
In the basis $\left\{ \ket{T_{11}},\ket{S_{11}},\ket{S_{02}}\right\} $
{[}Eqs. \eqref{eq:T11}-\eqref{eq:S02}{]}, $H_{\alpha}^{\prime}\equiv H_{\alpha}+H_{\alpha Z}$
takes the form \cite{Taylor2005,Taylor2007} 
\begin{equation}
H_{\alpha}^{\prime}=\left(\begin{array}{ccc}
0 & \omega_{\alpha Z} & 0\\
\omega_{\alpha Z} & 0 & \sqrt{2}t_{\alpha}\\
0 & \sqrt{2}t_{\alpha} & \Delta_{\alpha}
\end{array}\right),\label{eq:Halphaprime}
\end{equation}
where we have defined the energy associated with a static magnetic
field gradient of magnitude $dB_{\alpha}\equiv\left(B_{\alpha1}-B_{\alpha2}\right)/2$
along the single-spin quantization axis as $\omega_{\alpha Z}\equiv g\mu_{B}dB_{\alpha}.$ 

We initially consider the double charge resonant regime ($\Delta_{\alpha}\gg\delta$)
in the limit $\omega_{\alpha Z}\rightarrow0$ and keep only the term
$H_{\alpha}$ in the Hamiltonian for double dot $\alpha$. Elimination
of the doubly occupied singlet state $\ket{S_{02}}$ gives the effective
Hamiltonian $H_{\alpha J}\equiv-J_{\alpha}\ket{\tilde{S}_{11}}\bra{\tilde{S}_{11}}=J_{\alpha}\left(Z_{\alpha}-{\bf 1}\right)/2,$
where $Z\equiv\ket{T_{11}}\bra{T_{11}}-\ket{\tilde{S}_{11}}\bra{\tilde{S}_{11}},$
which generates a rotation around the $z$ axis of the Bloch sphere
for the singlet-triplet qubit \cite{Petta2005,Taylor2005,Taylor2007}.
Combining $H_{\alpha J}$ for $\alpha=a,b$ and $H_{{\rm eff}}$ {[}Eq.
\eqref{eq:Heff}{]} yields, in the two-qubit basis $\left\{ \ket{T_{11},T_{11}},\ket{\tilde{S}_{11},T_{11}},\ket{T_{11},\tilde{S}_{11}},\ket{\widetilde{S_{11},S_{11}}},\ket{\widetilde{S_{02},S_{02}}}\right\} ,$
\begin{equation}
H_{J}=\left(\begin{array}{ccc|cc}
0 &  & \\
 & -J_{a} & \\
 &  & -J_{b}\\
\hline  &  &  & -J_{a}-J_{b} & -j_{x}\\
 &  &  & -j_{x} & -J_{a}-J_{b}+j_{d}
\end{array}\right).\label{eq:HJ}
\end{equation}
The dynamics generated by $H_{J}$ are described by the operator
\begin{equation}
\hat{U}_{J}\left(\tau\right)\equiv e^{-iH_{J}\tau}=\left(\begin{array}{ccc|c}
1 &  & \\
 & e^{iJ_{a}\tau} & \\
 &  & e^{iJ_{b}\tau}\\
\hline  &  &  & e^{-iH_{{\rm eff}}\tau}
\end{array}\right),\label{eq:UJ}
\end{equation}
where
\begin{eqnarray}
e^{-iH_{{\rm eff}}\tau} & = & e^{i\left(J_{a}+J_{b}-j_{d}/2\right)\tau}\left[\cos\left(\frac{\Omega\tau}{2}\right)\mathbf{1}\right.\nonumber \\
 &  & \left.+i\sin\left(\frac{\Omega\tau}{2}\right)\left(\frac{j_{d}}{\Omega}\sigma_{z}+\frac{2j_{x}}{\Omega}\sigma_{x}\right)\right].\label{eq:expHeff}
\end{eqnarray}
The gate $\hat{U}_{J}$ thus describes an oscillation between $\ket{\widetilde{S_{11},S_{11}}}$
and $\ket{\widetilde{S_{02},S_{02}}}$ with frequency $\Omega,$ together
with $z$-axis rotations of the individual qubits. This evolution
occurs in the double charge resonant regime illustrated in Figs. \ref{fig:doubledouble}
and \ref{fig:energies}.

To obtain a controlled-phase gate using $\hat{U}_{J}$ that incorporates
robustness to single-qubit exchange errors, we construct a gate sequence
that includes spin-echo (refocusing) pulses \cite{Vandersypen2005}.
For a singlet-triplet qubit, phase errors accumulated due to exchange
fluctuations can be canceled via a $\pi$ rotation about the $x$
axis of the Bloch sphere \cite{Wu2002,Taylor2005}, and simultaneous
$\pi$ pulses can be applied to both qubits \cite{Shulman2012}. Since
single-qubit $x$-axis rotations are generated by the terms $H_{aZ}$
and $H_{bZ}$ {[}see Eqs. \eqref{eq:HalphaZ} and \eqref{eq:Halphaprime}{]},
the refocusing pulses are applied in the regime $\omega_{\alpha Z}\gg J_{\alpha}$
for $\alpha=a,b.$ This regime can be reached by adjusting the double-dot
detunings such that the $\ket{0211}$ and $\ket{1102}$ states are
energetically closer than $\ket{0202}$ to the $\ket{1111}$ state,
with $\Delta_{\alpha}\lesssim\left|\tilde{\delta}\right|-t_{\alpha}$
for $\alpha=a,b$ (here, we use a new symbol $\tilde{\delta}$ in
order to indicate that the range of values of $\Delta_{a}+\Delta_{b}-U_{ab}$
is different from that of $\delta$ in the double charge resonant
regime). The effective Hamiltonian is $H_{Z}\equiv\omega_{Z}\left(X_{a}+X_{b}\right)+\tilde{\delta}\ket{\widetilde{S_{02},S_{02}}}\bra{\widetilde{S_{02},S_{02}}},$
where $X\equiv\ket{T_{11}}\bra{\tilde{S}_{11}}+\ket{\tilde{S}_{11}}\bra{T_{11}}$
and we choose $dB_{a}=dB_{b}$ for simplicity. The associated evolution
is $e^{-iH_{Z}\tau},$ which for $\tau=\pi/2\omega_{Z}$ is equal
to $R_{\pi}\equiv-X_{a}X_{b}+e^{-i\pi\tilde{\delta}/2\omega_{Z}}\ket{\widetilde{S_{02},S_{02}}}\bra{\widetilde{S_{02},S_{02}}}.$
We note that applying $R_{\pi}$ results in the accumulation of a
relative phase between the $\ket{0202}$ and $\ket{1111}$ charge
subspaces. 

The full sequence for the controlled-phase gate in terms of the exchange
gate in the double charge resonant regime, $\hat{U}_{J},$ and the
refocusing pulse gate, $R_{\pi},$ is given by 
\begin{eqnarray}
 & P_{S}e^{-i\phi}e^{-iH_{aJ}\tau_{a}}e^{-iH_{bJ}\tau_{b}}R_{\pi}\hat{U}_{J}\left(\tau_{n}\right)R_{\pi}\hat{U}_{J}\left(\tau_{n}\right)P_{S}\nonumber \\
 & =\left(\begin{array}{cccc}
1\\
 & 1\\
 &  & 1\\
 &  &  & e^{2i\phi}
\end{array}\right).\label{eq:Ucphaseseq}
\end{eqnarray}
Here, $\tau_{n}=2\pi n/\Omega$ and $\phi=\left(1-j_{d}/\Omega\right)n\pi,$
where $n$ is an integer, $\tau_{\alpha}=\phi/J_{\alpha},$ and $P_{S}$
is the projector onto the four-dimensional $\ket{1111}$ subspace
spanned by $\left\{ \ket{T_{11},T_{11}},\ket{\tilde{S}_{11},T_{11}},\ket{T_{11},\tilde{S}_{11}},\ket{\widetilde{S_{11},S_{11}}}\right\} .$
In the next section, we consider the controlled $\pi$-phase gate,
which corresponds to $\phi=\pi/2.$

\section{Charge noise and gate fidelity\label{sec:ChgnoisegateF}}

In practice, the performance of the controlled-phase gate in Eq. \eqref{eq:Ucphaseseq}
is affected by charge noise \cite{Coish2005,Hu2006,Taylor2007,Meunier2011,Yang2011geometry,Dial2013}.
We now investigate the effects of classical, Gaussian-distributed
noise in $\delta$ and $\Delta_{d}$ due to gate voltage fluctuations
and set $\delta^{\prime}=\delta+\xi_{s},$ $\Delta_{d}^{\prime}=\Delta_{d}+\xi_{d},$
where $\xi_{s}$ and $\xi_{d}$ are assumed to be uncorrelated and
have the distributions $\text{\ensuremath{\rho_{\beta}\left(\xi_{\beta}\right)}}=e^{-\text{\ensuremath{\xi_{\beta}^{2}}}/2\text{\ensuremath{\sigma_{\beta}^{2}}}}/\sqrt{2\pi}\text{\ensuremath{\sigma_{\beta}}}$
with charge noise standard deviations $\sigma_{\beta}$ for $\beta=s,d.$
In what follows, we assume that $R_{\pi}$ and the single-qubit rotations
in Eq. \eqref{eq:Ucphaseseq} are ideal in order to focus on effects
due to errors in $\hat{U}_{J},$ which is the gate derived from the
capacitive interaction of the double dots in the double charge resonant
regime. Errors due to residual magnetic gradient terms are discussed
briefly at the end of this section. 

We therefore consider the simpler gate sequence 
\begin{equation}
U_{\phi}\equiv\hat{U}_{J}\left(\tau_{n}\right)R_{\pi}\hat{U}_{J}\left(\tau_{n}\right)=\left(\begin{array}{ccccc}
 &  &  & e^{i\phi}\\
 &  & 1\\
 & 1\\
e^{i\phi}\\
 &  &  &  & e^{i\zeta}
\end{array}\right),\label{eq:Uphi}
\end{equation}
where $\zeta=\left\{ 2\left(J_{a}+J_{b}-j_{d}\right)n/\Omega+1-\tilde{\delta}/2\omega_{Z}\right\} \pi$
and we have neglected a trivial global phase factor. Equation \eqref{eq:Uphi}
represents the ideal gate sequence. We determine the gate sequence
$U_{\phi}^{\prime}$ in the presence of charge noise by expanding
the terms in the Hamiltonian $H_{J}$ {[}Eq. \eqref{eq:HJ}{]}, which
are defined in Eqs. \eqref{eq:Ja}-\eqref{eq:jx}, up to second order
in the fluctuations $\xi_{\beta}.$ For $h=J_{a},J_{b},j_{d},j_{x},$
\begin{eqnarray*}
h^{\prime}\equiv h\left(\delta^{\prime},\Delta_{d}^{\prime}\right) & \approx & h\left(\delta,\Delta_{d}\right)+\left.\frac{\partial h}{\partial\delta^{\prime}}\right|_{0}\xi_{s}+\left.\frac{\partial h}{\partial\Delta_{d}^{\prime}}\right|_{0}\xi_{d}\\
 & + & \frac{1}{2}\left.\frac{\partial^{2}h}{\partial\delta^{\prime2}}\right|_{0}\xi_{s}^{2}+\frac{1}{2}\left.\frac{\partial^{2}h}{\partial\Delta_{d}^{\prime2}}\right|_{0}\xi_{d}^{2}
\end{eqnarray*}
 
\begin{eqnarray*}
 & + & \left.\frac{\partial^{2}h}{\partial\delta^{\prime}\partial\Delta_{d}^{\prime}}\right|_{0}\xi_{s}\xi_{d},
\end{eqnarray*}
where we use the notation $\left.\right|_{0}\equiv\left.\right|_{\delta^{\prime}=\delta,\Delta_{d}^{\prime}=\Delta_{d}}.$
Substitution of these expressions into Eq. \eqref{eq:UJ} then yields
$U_{\phi}^{\prime}.$ 

For an initial state $\ket{\psi_{in}},$ we define the minimum fidelity
as 
\begin{eqnarray}
F_{\min} & = & \left\langle {\rm Tr}\left[\hat{\rho}_{{\rm out}}^{\left(0\right)}\hat{\rho}_{{\rm out}}\right]\right\rangle _{\xi_{s},\xi_{d}}\nonumber \\
 & = & \int_{-\infty}^{\infty}\int_{-\infty}^{\infty}{\rm Tr}\left[\hat{\rho}_{{\rm out}}^{\left(0\right)}\hat{\rho}_{{\rm out}}\right]\nonumber \\
 &  & \ \ \ \ \ \ \ \ \ \ \ \times\rho_{s}\left(\xi_{s}\right)\rho_{d}\left(\xi_{d}\right)d\xi_{s}d\xi_{d},\label{eq:Fmin}
\end{eqnarray}
where $\hat{\rho}_{{\rm out}}^{\left(0\right)}\equiv U_{\phi}\ket{\psi_{{\rm in}}}\bra{\psi_{{\rm in}}}U_{\phi}^{\dagger}$
is the final state after evolution under the ideal gate sequence and
$\hat{\rho}_{{\rm out}}\equiv U_{\phi}^{\prime}\ket{\psi_{{\rm in}}}\bra{\psi_{{\rm in}}}U_{\phi}^{\prime\dagger}$
is the final state after the corresponding evolution in the presence
of charge noise. We choose $\ket{\psi_{{\rm in}}}=\frac{1}{2}\left(\ket{T_{11},T_{11}}+\ket{\tilde{S}_{11},T_{11}}+\ket{T_{11},\tilde{S}_{11}}+\ket{\widetilde{S_{11},S_{11}}}\right)$
in order to maximize the error (see Appendix \ref{sec:Minimum-Fidelity})
and assume that this state can be prepared without errors. $F_{{\rm min}}$
is then independent of $\zeta$, as there is initially zero probability
that the system is in the state $\ket{\widetilde{S_{02},S_{02}}}$
{[}see Eq. \eqref{eq:Uphi}{]}. We find
\begin{equation}
{\rm Tr}\left[\hat{\rho}_{{\rm out}}^{\left(0\right)}\hat{\rho}_{{\rm out}}\right]=\cos^{2}\left[\frac{n\pi}{2}\left(\frac{j_{d}^{\prime}}{\Omega^{\prime}}-\frac{j_{d}}{\Omega}\right)\right],\label{eq:Fminintegrand}
\end{equation}
where $\Omega^{\prime}=\sqrt{j_{d}^{\prime2}+4j_{x}^{\prime2}}.$ 

\begin{figure}
\includegraphics[bb=20bp 50bp 320bp 320bp,width=2.75in]{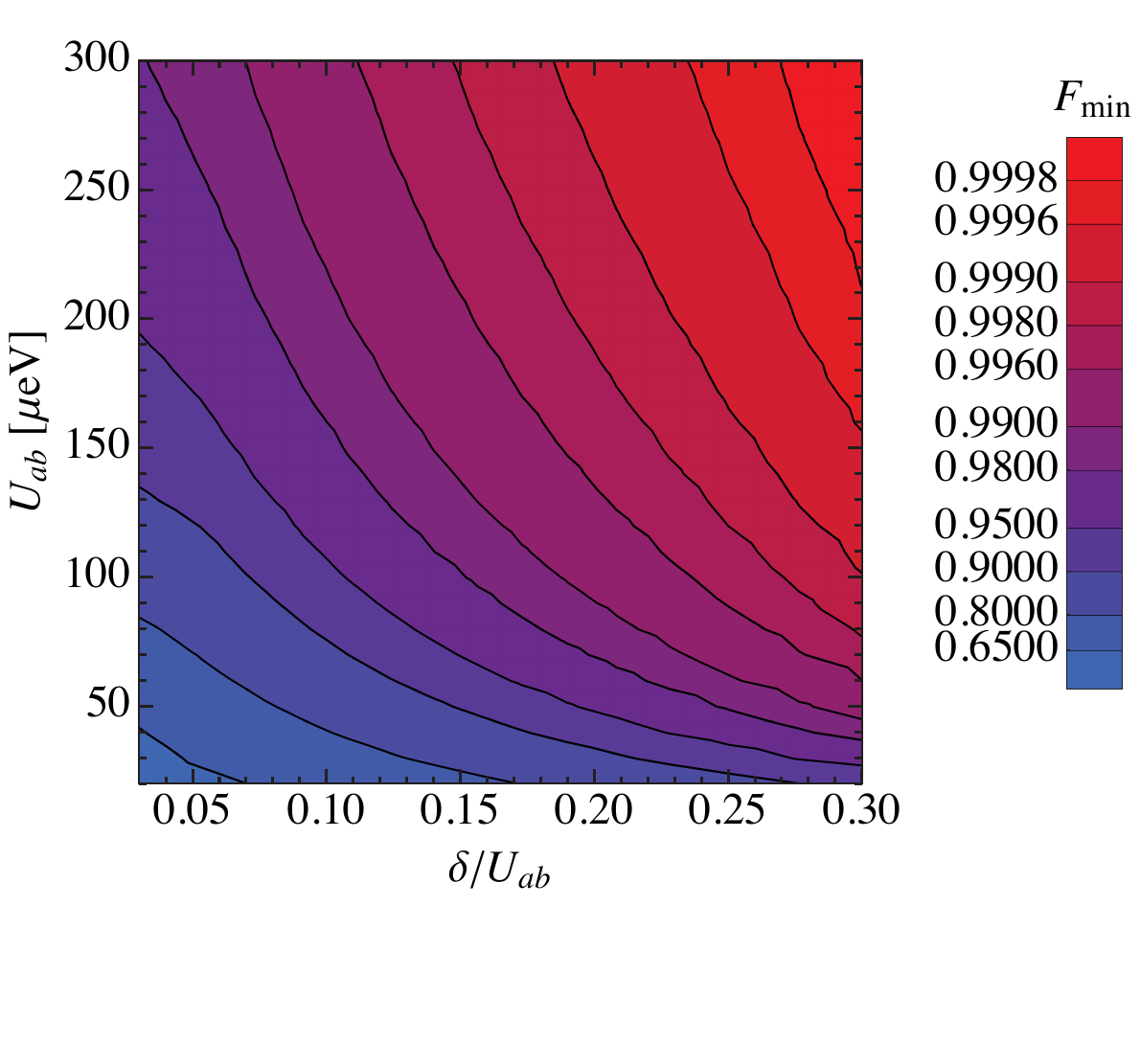}

\protect\caption{\label{fig:FminvsdeltaUab}Minimum fidelity $F_{{\rm min}}$ {[}Eq.
\eqref{eq:Fmin}{]} of the controlled-phase gate sequence {[}Eq. \eqref{eq:Uphi}{]}
for $\phi=\pi/2,$ $\Delta_{d}=0,$ and $\eta_{0}=0.1$ as a function
of the energy difference $\delta$ and the capacitive coupling strength
$U_{ab}$ {[}see Fig. \ref{fig:doubledouble}(b){]}. The values of
$\delta$ in the plot are calculated for each value of $U_{ab}$ by
solving the constraint $\phi=\pi/2$ with the chosen parameter values
for $n=1,2,\ldots,15$ (see the main text). Note that this implies
that the range of $\delta$ varies with $U_{ab}$ for fixed $\eta_{0}$.
The values of $\delta$ are therefore given in units of the value
of $U_{ab}$ with which they are associated.}
\end{figure}

We now calculate $F_{{\rm min}}$ for the controlled $\pi$-phase
gate. The associated constraint $\phi=\pi/2$ {[}see Eq. \eqref{eq:Ucphaseseq}{]}
leads to $n=\left[2\left(1-j_{d}/\Omega\right)\right]^{-1}.$ Since
$n$ must be an integer, this relation restricts the possible values
of $\delta$ and $\Delta_{d}$ for fixed values of the charge admixture
parameters $\eta_{a},$ $\eta_{b}$ and the capacitive coupling strength
$U_{ab}.$ For the parameter regime we consider in the present work,
we find that $n$ varies more strongly with $\delta$ than with $\Delta_{d}$
and set $\Delta_{d}=0$ for simplicity in the remainder of the analysis.
Choosing $\eta_{a}=\eta_{b}\equiv\eta_{0}=0.1$ and $\sigma_{s}=2\ \mu{\rm eV}$
\cite{Petersson2010}, we solve the constraint for the values of $\delta$
corresponding to $n=1,2,\ldots,15$ and calculate $F_{\min}$ via
numerical integration using Eq. \eqref{eq:Fmin}. By repeating this
calculation for a range of coupling strengths $U_{ab},$ we obtain
the variation of $F_{\min}$ with $\delta$ and $U_{ab}$ shown in
Fig. \ref{fig:FminvsdeltaUab}. We see that $F_{{\rm min}}$ increases
with increasing coupling strength $U_{ab}$ as expected. For a given
value of $U_{ab},$ $F_{{\rm min}}$ also increases as $\delta$ increases,
i.e., as the energy separation between $\ket{S_{11},S_{11}}$ and
$\ket{S_{02},S_{02}}$ becomes larger and the contribution of $\ket{S_{02},S_{02}}$
to the ground state decreases. For $\delta=41\ \mu{\rm eV}$ (corresponding
to $n=14$) and $U_{ab}=150\ \mu{\rm eV},$ $F_{\min}>0.999$ and
the time for the gate $\hat{U}_{J}$ is $\tau_{n}=2\pi n/\Omega\approx1\ {\rm ns.}$ 

While we assume ideal echo pulses $R_{\pi}$ in the present analysis,
switching times for the magnetic gradients used to generate $R_{\pi}$
which are longer than the timescale of the qubit dynamics will lead
to residual magnetic gradients that remain during the action of the
exchange gate $\hat{U}_{J}$. The residual Zeeman energy $\omega_{Z,{\rm res}}$
will modify the Hamiltonian in Eq. \eqref{eq:HJ} and thus lead to
errors. Setting $J_{a}=J_{b}\equiv J_{0}$ and assuming $\omega_{Z,{\rm res}}\ll J_{0},$
we can regard the residual magnetic gradient terms as a perturbation
to $H_{J}$ and obtain an upper bound for the allowable residual gradient
$dB_{{\rm res}}$ using the condition $\tau_{n}\omega_{Z,{\rm res}}^{2}/J_{0}\ll1.$
For $\eta_{0}=0.1,$ $\delta=41\ \mu{\rm eV},$ and $U_{ab}=150\ \mu{\rm eV},$
this condition yields $\omega_{Z,{\rm res}}\ll1\ \mu{\rm eV},$ corresponding
to $dB_{{\rm res}}\ll40\ {\rm mT}$ for GaAs dots (with $g=-0.44$)
and $dB_{{\rm res}}\ll10\ {\rm mT}$ for Si dots (with $g=2$). 

\section{Charge relaxation via phonons \label{sec:phononrelaxation}}

\subsection{Relaxation rate}

In addition to charge noise arising from gate voltage fluctuations,
charge relaxation due to electron-phonon coupling also affects the
coherence of the capacitively coupled double-dot system we consider.
Here, we determine the rate $\Gamma$ of relaxation via phonons that
occurs between the eigenstates $\ket{g}$ and $\ket{e}$ of $H_{{\rm eff}}$
{[}Eq. \eqref{eq:Heff}{]}, as illustrated in Fig. \ref{fig:energies}.
From Fermi's golden rule, $\Gamma\sim\left|\left\langle g\right|H_{\mathrm{ep}}\left|e\right\rangle \right|^{2}\rho\left(\Omega\right),$
where $H_{{\rm ep}}$ is the electron-phonon interaction Hamiltonian
and $\rho\left(\Omega\right)$ is the phonon density of states at
the gap energy $\Omega.$ In the following analysis, we consider acoustic
phonons in both GaAs and Si quantum dots and calculate $\Gamma$ for
the higher-energy state, assuming $k_{B}T\lesssim\hbar j_{x}$. The
electron-phonon interaction for GaAs is described by the Hamiltonian
\cite{Mahan2000}
\begin{eqnarray}
H_{\mathrm{GaAs}} & = & \sum_{\mu,\mathbf{k}}\sqrt{\frac{\hbar}{2\rho_{0}V_{0}c_{\mu}k}}\left(k\Xi{}_{l}\delta_{\mu,l}-i\beta\right)\nonumber \\
 &  & \ \ \ \ \ \ \ \ \times\left(a_{\mu,\mathbf{k}}+\ a_{\mu,-\mathbf{k}}^{\dagger}\right)M_{k},\label{eq:HepGaAs}
\end{eqnarray}
while the Hamiltonian for Si has the form \cite{Yu2010} 
\begin{eqnarray}
H_{\mathrm{Si}} & = & i\sum_{\mu,{\bf k}}\sqrt{\frac{\hbar}{2\rho_{0}V_{0}c_{\mu}k}}\left({\bf k}\cdot\hat{\boldsymbol{\epsilon}}_{\mu,{\bf k}}\ \Xi_{d}\right.\nonumber \\
 &  & \ \ \left.+\ k_{z^{\prime}}\,\hat{z}^{\prime}\cdot\hat{\boldsymbol{\epsilon}}_{\mu,{\bf k}}\ \Xi_{u}\right)\left(a_{\mu,{\bf k}}+a_{\mu,-{\bf k}}^{\dagger}\right)M_{k}.\label{eq:HepSi}
\end{eqnarray}
In Eqs. \eqref{eq:HepGaAs} and \eqref{eq:HepSi}, $a_{\mu,\mathbf{k}}^{\dagger}$
creates an acoustic phonon with wave vector $\mathbf{k}$, polarization
$\mu$ {[}the sum is taken over one longitudinal mode ($\mu=l$) and
two transverse modes ($\mu=p$){]}, phonon speed $c_{\mu},$ energy
$\varepsilon_{\mathrm{ph}}=\hbar c_{\mu}k,$ and unit polarization
vector $\hat{\boldsymbol{\epsilon}}_{\mu,{\bf k}},$ $\rho_{0}$ is
the mass density of the material, $V_{0}$ is the crystal volume,
$\Xi_{l}$ is the deformation potential and $\beta$ is the piezoelectric
constant for GaAs, $\Xi_{d}$ ($\Xi_{u}$) is the dilation (uniaxial)
deformation potential for Si, $\hat{z}^{\prime}$ denotes the direction
of uniaxial strain, and $\delta_{\mu,l}$ is the Kronecker delta function.
The different phonon terms appearing in Eqs. \eqref{eq:HepGaAs} and
\eqref{eq:HepSi} reflect the fact that the crystal structure of GaAs
lacks a center of symmetry, whereas unstrained Si has a centrosymmetric
crystal structure: while both deformation potential and piezoelectric
phonons contribute to the electron-phonon coupling in GaAs, there
is no contribution from piezoelectric phonons for Si \cite{Mahan2000}.
Thus, the strength of the electron-phonon coupling and the associated
relaxation rate are expected to be much smaller for Si quantum dots
\cite{Zwanenburg2013}. 

\begin{figure*}
\includegraphics[bb=50bp 30bp 570bp 460bp,width=0.6\paperwidth]{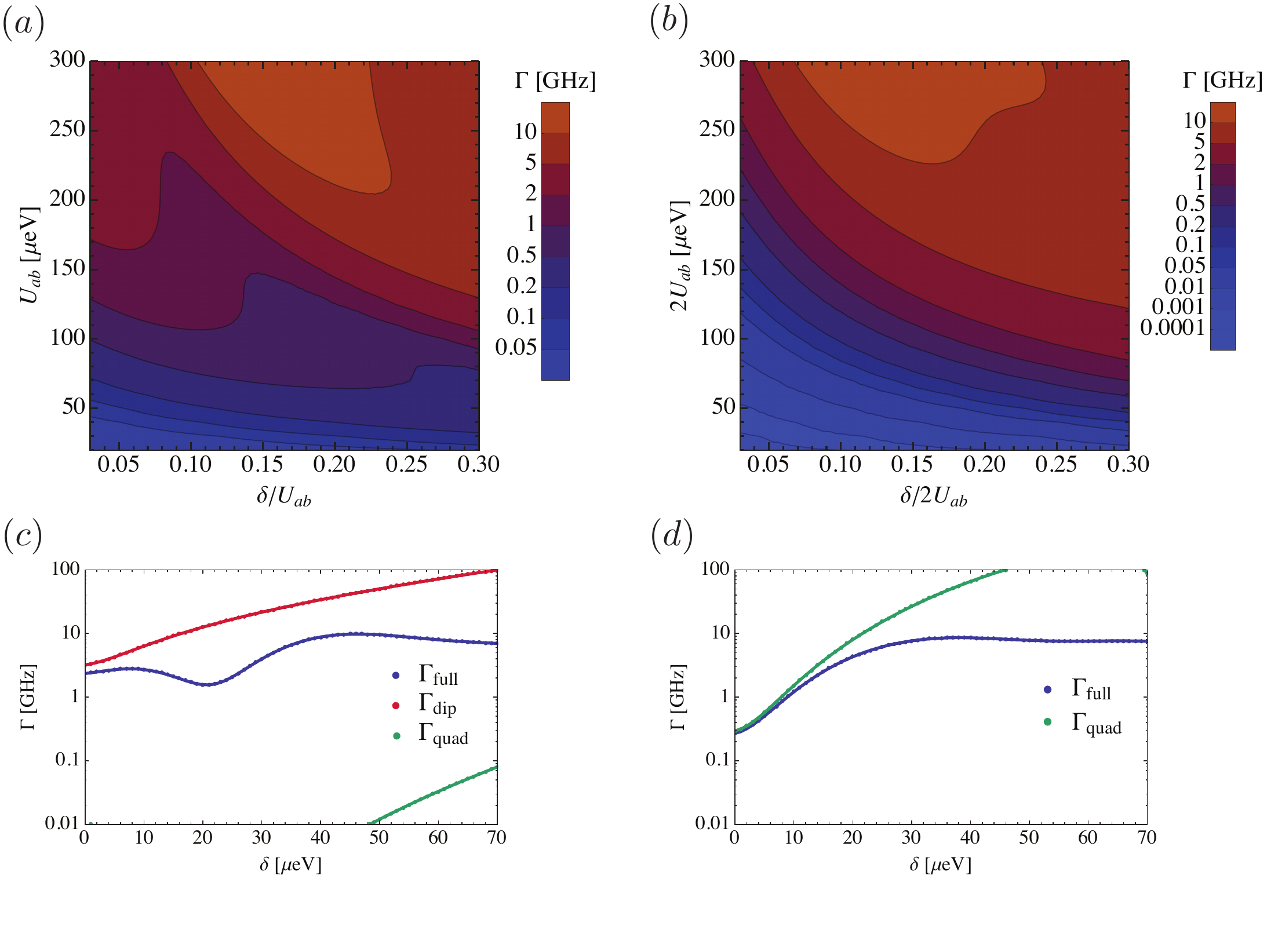}

\protect\caption{\label{fig:phononGaAs}Rate of relaxation via electron-phonon coupling
for capacitively coupled GaAs double quantum dots {[}Eq. \eqref{eq:HepGaAs}{]}.
The rate is calculated as a function of $\delta$ and the total interqubit
capacitive coupling strength {[}see Fig. \ref{fig:doubledouble}(b){]},
which is equal to $U_{ab}$ for the linear dot geometry (a) and $2U_{ab}$
for the purely quadrupolar dot geometry (b) considered in the present
work (see the main text). The parameter values used are $\Delta_{d}=0,$
$\eta_{0}=0.1,$ $d=140\ {\rm nm},$ $R=2d,$ the GaAs effective mass
$m^{\ast}=0.067m_{e}$ (where $m_{e}$ is the free-electron mass),
dot size $\sigma=20\ {\rm nm,}$ and phonon parameter values $\rho_{0}=5.3\times10^{3}\ \mbox{kg/m}^{3},$
$c_{l}=5.3\times10^{3}\ \mathrm{m}/\mathrm{s},$ $c_{t}=2.5\times10^{3}\ \mathrm{m}/\mathrm{s},$
$\Xi_{l}=7\ \mbox{eV},$ and $\beta=1.4\times10^{9}\ \mbox{eV/m}$
\cite{Stano2006PRL}. (c) Dipolar ($\Gamma_{{\rm dip}}$) and quadrupolar
($\Gamma_{{\rm quad}}$) contributions to the full relaxation rate
($\Gamma_{{\rm full}}$) for the linear geometry of Fig. \ref{fig:doubledouble}(a)
with $U_{ab}=200\ \mu{\rm eV}.$ (d) Comparison of $\Gamma_{{\rm quad}}$
and $\Gamma_{{\rm full}}$ for the purely quadrupolar geometry, corresponding
to $\theta=\pi/2$ and $\varphi=0$ in Fig. \ref{fig:ccgeometry}(a),
with $2U_{ab}=200\ \mu{\rm eV}.$}
\end{figure*}

\begin{figure*}
\includegraphics[bb=50bp 30bp 570bp 460bp,width=0.6\paperwidth]{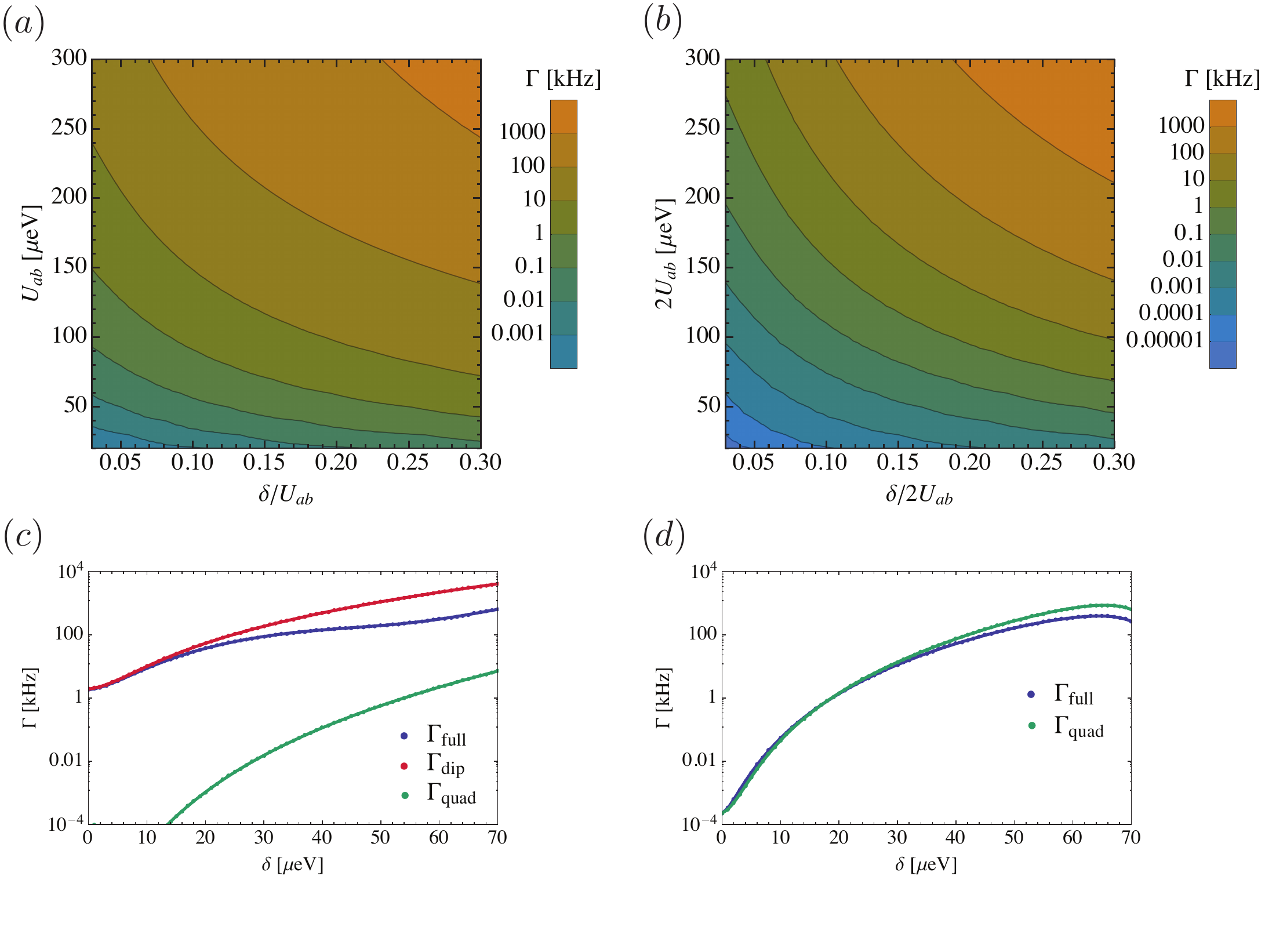}

\protect\caption{\label{fig:phononSi}Rate of relaxation via electron-phonon coupling
for capacitively coupled Si double quantum dots {[}Eq. \eqref{eq:HepSi}{]}.
The rate is calculated as a function of $\delta$ and the total interqubit
capacitive coupling strength {[}see Fig. \ref{fig:doubledouble}(b){]},
which is equal to $U_{ab}$ for the linear dot geometry (a) and $2U_{ab}$
for the purely quadrupolar dot geometry (b) considered in the present
work (see the main text). The parameter values used are $\Delta_{d}=0,$
$\eta_{0}=0.1,$ $d=140\ {\rm nm},$ $R=2d,$ the Si effective mass
$m^{\ast}=0.19m_{e},$ dot size $\sigma=22\ {\rm nm,}$ and phonon
parameter values $\rho_{0}=2.33\times10^{3}\ \mbox{kg/m}^{3},$ $c_{l}=9.33\times10^{3}\ \mathrm{m}/\mathrm{s},$
$c_{t}=5.42\times10^{3}\ \mathrm{m}/\mathrm{s},$ $\Xi_{d}=5\ \mbox{eV},$
and $\Xi_{u}=8.77\ \mbox{eV}$ \cite{Yu2010,Tahan2014}. (c) Dipolar
($\Gamma_{{\rm dip}}$) and quadrupolar ($\Gamma_{{\rm quad}}$) contributions
to the full relaxation rate ($\Gamma_{{\rm full}}$) for the linear
geometry of Fig. \ref{fig:doubledouble}(a) with $U_{ab}=200\ \mu{\rm eV}.$
(d) Comparison of $\Gamma_{{\rm quad}}$ and $\Gamma_{{\rm full}}$
for the purely quadrupolar geometry, corresponding to $\theta=\pi/2$
and $\varphi=0$ in Fig. \ref{fig:ccgeometry}(a), with $2U_{ab}=200\ \mu{\rm eV}.$}
\end{figure*}

The factor in $H_{{\rm GaAs}}$ and $H_{{\rm Si}}$ encompassing the
coupling to electron charge degrees of freedom is $M_{k}=M_{k}^{\left(a\right)}+M_{k}^{\left(b\right)},$
where
\begin{equation}
M_{k}^{\left(\alpha\right)}\equiv\sum_{i,j=1,2}\sum_{\sigma}\bra{\alpha,i}e^{i\mathbf{k}\cdot\mathbf{r}}\ket{\alpha,j}c_{\alpha i\sigma}^{\text{\dag}}c_{\alpha j\sigma}\label{eq:Mkalpha}
\end{equation}
and ${\bf r}$ is the electron position operator. Note that in our
calculation, we take each double dot to be coupled independently to
the same phonon bath \cite{Brandes2002,Barrett2002}. Thus, we implicitly
assume that the phonon mean free path is greater than the size of
the system, so that scattering of phonons between interactions with
the electron pairs in the two double dots can be neglected. The matrix
elements in Eq. \eqref{eq:Mkalpha} depend on the spatial configuration
of the four quantum dots and are evaluated using the two-dimensional
Gaussian wave functions $\Psi_{\alpha i}\left({\bf r}\right)\equiv\langle{\bf r}\ket{\alpha,i}=\psi\left(x-x_{\alpha i}\right)\psi\left(y-y_{\alpha i}\right)$
for $i=1,2$ and $\alpha=a,b,$ where $\psi\left(q\right)=e^{-q^{2}/4\sigma^{2}}/\left(2\pi\sigma^{2}\right)^{1/4}.$
Re-expressing $M_{k}^{\left(a\right)}$ and $M_{k}^{\left(b\right)}$
in the basis $\left\{ \ket{S_{11},S_{11}},\ket{S_{02},S_{02}},\ket{S_{11},S_{02}},\ket{S_{02},S_{11}}\right\} $
and using the same Schrieffer-Wolff transformation used for $H_{{\rm hub}}$
in Sec. \ref{sec:Model} to write $\tilde{M}_{k}=e^{\lambda A}M_{k}e^{-\lambda A}\approx M_{k}+\lambda\left[A,M_{k}\right]+\frac{\lambda^{2}}{2}\left[A,\left[A,M_{k}\right]\right],$
we determine the transition matrix element 
\begin{eqnarray*}
\bra{g}M_{k}\ket{e} & = & \cos\theta\sin\theta\left(\bra{S_{11},S_{11}}\tilde{M}_{k}\ket{S_{11},S_{11}}\right.\\
 &  & \left.-\bra{S_{02},S_{02}}\tilde{M}_{k}\ket{S_{02},S_{02}}\right)\\
 &  & +\cos^{2}\theta\bra{S_{11},S_{11}}\tilde{M}_{k}\ket{S_{02},S_{02}}\\
 &  & -\sin^{2}\theta\bra{S_{02},S_{02}}\tilde{M}_{k}\ket{S_{11},S_{11}}.
\end{eqnarray*}

The relaxation rates for GaAs and Si are given by $\Gamma_{{\rm GaAs}}=g_{l}I\left(\Omega/\hbar c_{l}\right)+g_{p}I\left(\Omega/\hbar c_{p}\right)$
and $\Gamma_{{\rm Si}}=s_{l}K_{l}\left(\Omega/\hbar c_{l}\right)+s_{p}K_{p}\left(\Omega/\hbar c_{p}\right),$
with the momentum-space angular integrals 
\begin{eqnarray}
I\left(k\right) & \equiv & \int\left|\bra{g}M_{k}\ket{e}\right|^{2}d\Omega_{{\rm ang}},\label{eq:Ik}\\
K_{l}\left(k\right) & \equiv & \int\left(1+\gamma\cos^{2}\chi\right)^{2}\left|\bra{g}M_{k}\ket{e}\right|^{2}d\Omega_{{\rm ang}},\label{eq:Klk}\\
K_{p}\left(k\right) & \equiv & \int\gamma^{2}\cos^{2}\chi\sin^{2}\chi\left|\bra{g}M_{k}\ket{e}\right|^{2}d\Omega_{{\rm ang}}\label{eq:Kpk}
\end{eqnarray}
and the factors
\begin{eqnarray}
g_{l} & = & \frac{\Omega}{8\pi^{2}\hbar^{2}\rho_{0}c_{l}^{3}}\left(\frac{\Omega^{2}}{\hbar^{2}c_{l}^{2}}\Xi_{l}^{2}+\beta^{2}\right),\label{eq:gl}
\end{eqnarray}
 
\begin{eqnarray}
g_{p} & = & \frac{2\Omega}{8\pi^{2}\hbar^{2}\rho_{0}c_{p}^{3}}\beta^{2},\label{eq:gp}\\
s_{\mu} & = & \frac{\Omega^{3}}{8\pi^{2}\hbar^{4}\rho_{0}c_{\mu}^{5}}\Xi_{d}^{2},\ \ \mu=l,p.\label{eq:smu}
\end{eqnarray}
In writing Eqs. \eqref{eq:Klk} and \eqref{eq:Kpk}, we have chosen
one of the two transverse ($\mu=p$) phonon polarization axes to lie
orthogonal to $\hat{z}^{\prime}$ {[}see Eq. \eqref{eq:HepSi}{]}
and defined $\chi$ as the angle between ${\bf k}$ and $\hat{z}^{\prime}.$
We also define $\gamma\equiv\Xi_{u}/\Xi_{d},$ and $\Omega_{{\rm ang}}$
denotes the momentum-space solid angle. 

We calculate the relaxation rates via numerical integration for the
linear geometry depicted in Fig. \ref{fig:doubledouble}(a), which
has both dipolar and quadrupolar moments, as well as for a purely
quadrupolar geometry, which corresponds to a rectangular arrangement
of the dots obtained from the general configuration illustrated in
Fig. \ref{fig:ccgeometry}(a) by setting $\theta=\pi/2$ and $\varphi=0.$
For the linear case (corresponding to $\theta=0,$ $\varphi=\pi$),
we set $x_{a1}=-\left(R+d\right)/2,$ $x_{a2}=-\left(R-d\right)/2,$
$x_{b1}=\left(R-d\right)/2,$ $x_{b2}=\left(R+d\right)/2,$ and $y_{\alpha i}=0$
for all $\alpha$ and $i.$ The coordinates of the dot centers for
the pure quadrupole are $\left(x_{a1},y_{a1}\right)=\left(-d/2,R/2\right),$
$\left(x_{a2},y_{a2}\right)=\left(d/2,R/2\right),$ $\left(x_{b1},y_{b1}\right)=\left(d/2,-R/2\right),$
$\left(x_{b2},y_{b2}\right)=\left(-d/2,-R/2\right),$ and we take
as the interqubit Coulomb interaction term for the quadrupolar geometry
\begin{eqnarray}
H_{{\rm int}} & = & U_{ab}\left(n_{a2}n_{b1}+n_{a1}n_{b2}\right).\label{eq:Hintquad}
\end{eqnarray}
Equation \eqref{eq:Hintquad} leads to $\delta=\Delta_{a}+\Delta_{b}-2U_{ab}$
and corresponding modifications to Eqs. \eqref{eq:Ja}-\eqref{eq:jx}
for the case of the purely quadrupolar system. Note that the total
coupling strength between the qubits for the quadrupolar geometry
is effectively twice that for the linear geometry. We therefore vary
$2U_{ab}$ for the quadrupolar system over the same range of values
of $U_{ab}$ considered for the linear configuration, in order to
focus on the geometry-dependent variation in the relaxation rate. 

The calculated relaxation rates are shown for GaAs in Fig. \ref{fig:phononGaAs}
and for Si in Fig. \ref{fig:phononSi} as a function of $\delta$
and the interqubit capacitive coupling strength, where we choose $\Delta_{d}=0,$
$\eta_{0}=0.1,$ $d=140\ {\rm nm},$ and $R=2d.$ Comparing Figs.
\ref{fig:phononGaAs}(a) for the linear geometry and \ref{fig:phononGaAs}(b)
for the purely quadrupolar geometry, we see that both relaxation rates
increase with increasing $U_{ab}$ but exhibit a nonmonotonic dependence
on $\delta.$ While the largest rates shown for both geometries are
$\sim10\ {\rm GHz},$ the rate for the quadrupolar geometry reduces
to $\lesssim100\ {\rm kHz}$ for the smallest values of $\delta$
and $U_{ab}$ considered. On the other hand, we see from Fig. \ref{fig:phononGaAs}(a)
that the rate reduces only to $\sim10\ {\rm MHz}$ for the linear
geometry. Figures \ref{fig:phononSi}(a) and \ref{fig:phononSi}(b)
reveal that the relaxation rates for Si dots are several orders of
magnitude smaller than those for GaAs dots, as expected due to the
absence of piezoelectric phonons in Si \cite{Zwanenburg2013}. The
rates increase as both $U_{ab}$ and $\delta$ are increased, with
a maximum rate $\sim1\ {\rm MHz}$ for the parameter ranges considered.
At the smallest values of $\delta$ and $U_{ab}$ shown, relaxation
for the linear geometry has a rate $\sim1\ {\rm Hz}$ {[}Fig. \ref{fig:phononSi}(a){]},
while the rate for the purely quadrupolar geometry {[}Fig. \ref{fig:phononSi}(b){]}
is two orders of magnitude smaller. 

We now consider separately the contributions of the dipolar and quadrupolar
terms in $M_{k}$ to the total relaxation rates for coupled GaAs and
Si double dots in both the linear and the purely quadrupolar geometries.
For phonon wavelengths long compared to the size of the quantum dot
system, we can write $e^{i{\bf k}\cdot{\bf r}}\approx1+i{\bf k}\cdot{\bf r}-\left({\bf k}\cdot{\bf r}\right)^{2}/2.$
The dipolar ($\Gamma_{{\rm dip}}$) and quadrupolar ($\Gamma_{{\rm quad}}$)
contributions to the rate are then obtained by calculating the relaxation
rates with the transition matrix elements $\bra{g}i{\bf k}\cdot{\bf r}\ket{e}$
and $\bra{g}\left({\bf k}\cdot{\bf r}\right)^{2}/2\ket{e},$ respectively,
substituted for the full matrix element $\bra{g}M_{k}\ket{e}$ in
Eqs. \eqref{eq:Ik}-\eqref{eq:Kpk}. We see in Figs. \ref{fig:phononGaAs}(c)
and \ref{fig:phononSi}(c) that, for both GaAs and Si, the full relaxation
rate $\Gamma_{{\rm full}}$ for the linear geometry contains a large
dipolar contribution and a much smaller quadrupolar contribution.
The large dipolar term can be understood from the fact that a net
dipole moment exists for the four-electron system in the linear configuration.
In contrast, the purely quadrupolar geometry {[}Figs. \ref{fig:phononGaAs}(d)
and \ref{fig:phononSi}(d){]} lacks a net dipole moment, so that $\Gamma_{{\rm dip}}=0$
in this case. While a large discrepancy exists between the quadrupolar
contribution $\Gamma_{{\rm quad}}$ and $\Gamma_{{\rm full}}$ for
GaAs, $\Gamma_{{\rm full}}$ for Si is well described by the quadrupolar
term. This can be understood from the fact that, over the range of
$\delta$ (and thus $\Omega$) we consider, the ratio of the system
size ($\sim R$) to the phonon wavelength is less than 1 for Si. On
the other hand, the corresponding ratio for GaAs becomes larger than
1 at sufficiently large values of $\delta.$

\subsection{Modification of controlled-Z gate fidelity \label{sub:ModCZfidelity}}

Having calculated the rate of phonon-induced charge relaxation within
the two-singlet subspace spanned by $\left\{ \ket{\widetilde{S_{11},S_{11}}},\ket{\widetilde{S_{02},S_{02}}}\right\} $,
we now determine the effect of this decay on the gate fidelity calculated
in Sec. \ref{sec:ChgnoisegateF} for the linear quantum dot geometry.
In order to incorporate the relaxation into the dynamics, we consider
the Lindblad master equation for the density matrix within the two-singlet
subspace $\hat{\rho}_{s}$, which can be written in the form
\begin{equation}
\dot{\hat{\rho}}_{s}=-i\left[\tilde{H},\hat{\rho}_{s}\right]+\Gamma a\hat{\rho}_{s}a^{\dagger}\label{eq:mastereq}
\end{equation}
with $\hat{a}\equiv\ket{\widetilde{S_{11},S_{11}}}\bra{\widetilde{S_{02},S_{02}}}$
and 
\begin{align}
\tilde{H} & \equiv H_{{\rm eff}}-i\frac{\Gamma}{2}\hat{a}^{\dagger}\hat{a}=H_{{\rm eff}}-i\frac{\Gamma}{2}\ket{\widetilde{S_{02},S_{02}}}\bra{\widetilde{S_{02},S_{02}}}.\label{eq:Htilde}
\end{align}
We assume $\Gamma,j_{x}\ll j_{d}$ and neglect the final (quantum-jump)
term in Eq. \eqref{eq:mastereq}. Within this approximation, we can
regard the dynamics in the two-singlet subspace {[}Eq. \eqref{eq:expHeff}{]}
as being generated by the non-Hermitian ``Hamiltonian'' in Eq. \eqref{eq:Htilde}
instead of $H_{{\rm eff}}.$ 

We can then estimate the effect of the phonon decay on the dynamics
by making the replacements 
\begin{eqnarray*}
e^{i\left[(J_{a}+J_{b})\tau_{n}+\phi\right]}\ket{\widetilde{S_{11},S_{11}}}\bra{\widetilde{S_{11},S_{11}}}\ \ \ \ \ \ \ \ \ \ \ \ \ \ \ \ \ \\
\ \ \ \ \rightarrow e^{-\Gamma_{{\rm eff}}\tau_{n}}e^{i\left[(J_{a}+J_{b})\tau_{n}+\phi\right]}\ket{\widetilde{S_{11},S_{11}}}\bra{\widetilde{S_{11},S_{11}}},\\
e^{i\left[(J_{a}+J_{b})\tau_{n}+\phi\right]}\ket{\widetilde{S_{02},S_{02}}}\bra{\widetilde{S_{02},S_{02}}}\ \ \ \ \ \ \ \ \ \ \ \ \ \ \ \ \ \\
\rightarrow e^{-\Gamma_{{\rm 2}}\tau_{n}}e^{i\left[(J_{a}+J_{b})\tau_{n}+\phi\right]}\ket{\widetilde{S_{02},S_{02}}}\bra{\widetilde{S_{02},S_{02}}}
\end{eqnarray*}
in $\hat{U}_{J}\left(\tau_{n}\right)$ {[}see Eq. \eqref{eq:UJ}{]}.
Here, $\Gamma_{{\rm {\rm eff}}}\equiv\Gamma j_{x}^{2}/2j_{d}^{2}$
and the explicit form of $\Gamma_{2}$ does not enter into the calculation
for our choice of initial state $\ket{\psi_{{\rm in}}},$ which is
defined entirely within the effective $\ket{1111}$ subspace. Incorporating
these modifications into the gate sequence in Eq. \eqref{eq:Uphi},
we determine the resulting modified minimum gate fidelity $F_{\min}^{\prime}$
using Eq. \eqref{eq:Fmin}. 

\begin{figure*}
\includegraphics[bb=50bp 30bp 560bp 810bp,width=0.5\paperwidth]{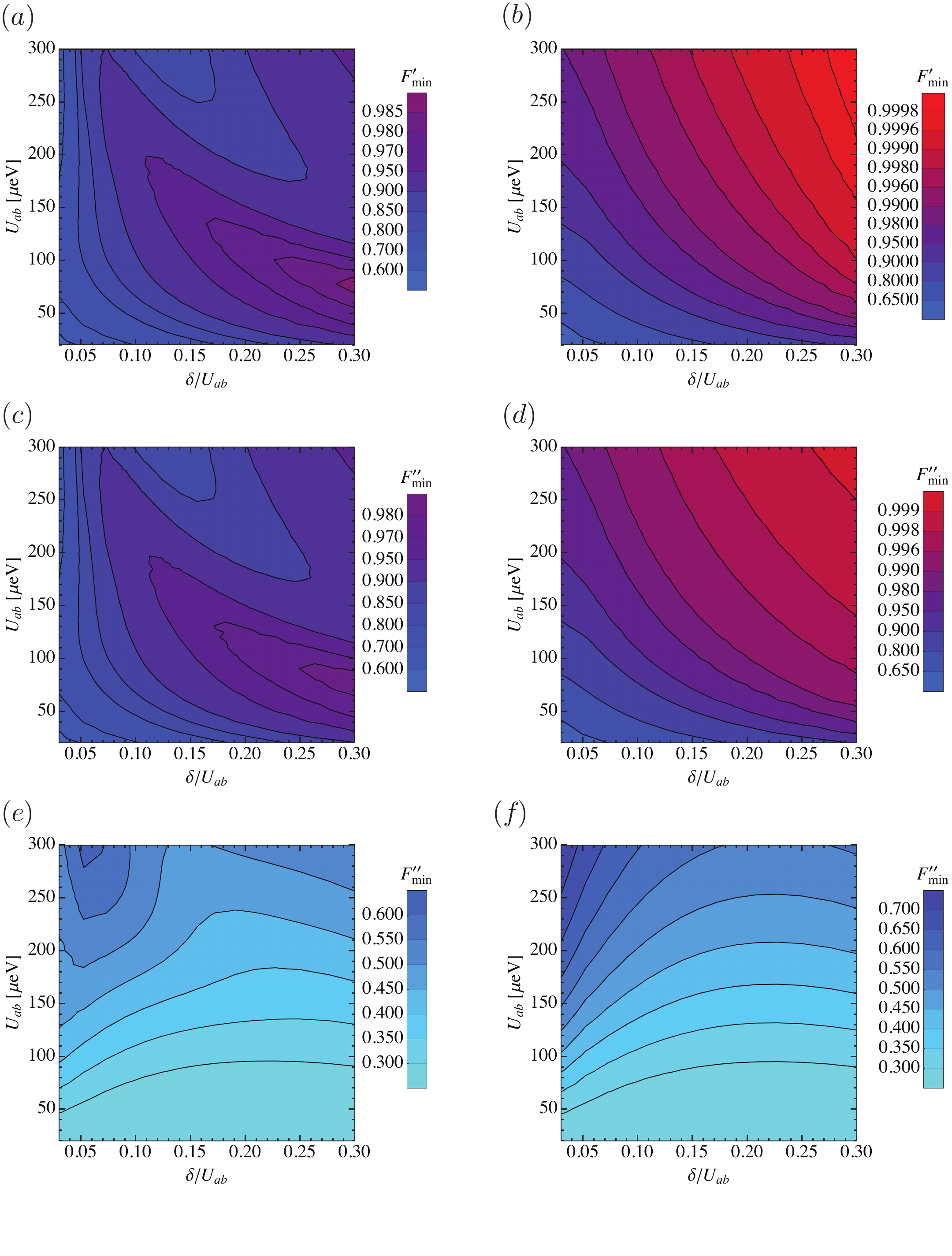}

\protect\caption{\label{fig:FminprimevsdeltaUab}(a),(b) Modified minimum fidelity
$F_{\min}^{\prime}$ of the controlled-phase gate sequence {[}Eq.
\eqref{eq:Uphi}{]} for $\phi=\pi/2$, in the presence of decay due
to electron-phonon coupling in (a) GaAs and (b) Si quantum dots arranged
in the linear geometry of Fig. \ref{fig:doubledouble}(a). The parameters
used in the calculation are identical to those given in the captions
of Figs. \ref{fig:FminvsdeltaUab},\ref{fig:phononGaAs}, and \ref{fig:phononSi}.
(c)-(f) Modified minimum fidelity $F_{{\rm \min}}^{\prime\prime}$
in the presence of both phonon-induced decay and charge noise of frequency
$\Gamma_{{\rm chg}}$, calculated for (c) GaAs dots with $\Gamma_{{\rm chg}}=1\ {\rm MHz},$
(d) Si dots with $\Gamma_{{\rm chg}}=1\ {\rm MHz},$ (e) GaAs dots
with $\Gamma_{{\rm chg}}=1\ {\rm GHz},$ and (f) Si dots with $\Gamma_{{\rm chg}}=1\ {\rm GHz}.$ }
\end{figure*}

The results are shown in Fig. \ref{fig:FminprimevsdeltaUab}(a) for
GaAs dots and in Fig. \ref{fig:FminprimevsdeltaUab}(b) for Si dots.
Comparing these plots with Fig. \ref{fig:FminvsdeltaUab}, we see
that the phonon-induced decay results in a large reduction of the
gate fidelity for GaAs, as expected from the fact that $\Gamma_{{\rm GaAs}}\sim100\ {\rm MHz}-10\ {\rm GHz}$
is comparable to $1/\tau_{n}.$ In contrast, essentially no modification
to the fidelity occurs for the case of Si, since $\Gamma_{{\rm Si}}\lesssim1\ {\rm MHz}\ll1/\tau_{n}.$
Thus, we find that an implementation of the controlled-Z gate based
on Si quantum dots provides robustness to phonon-induced decay. 

While the analysis so far has assumed quasistatic charge noise, fast
charge noise also affects the coherence of singlet-triplet spin qubits
in practice \cite{Dial2013}. Accordingly, we also calculate the modified
minimum fidelity $F_{\min}^{\prime\prime}$ of the controlled-phase
gate in the presence of decay with an effective rate $\Gamma_{{\rm eff}}^{\prime}\equiv\Gamma_{{\rm eff}}+\Gamma_{{\rm chg}},$
where $\Gamma_{{\rm chg}}$ is the fast charge noise frequency. Note
that, for realistic devices, $\Gamma_{{\rm chg}}$ may vary with the
particular operating point via the dependence of the detuning noise
spectral density on this point \cite{Dial2013}. Here, we assume that
$\Gamma_{{\rm chg}}$ is independent of the operating point for simplicity.
The results for GaAs (Si) dots are shown for $\Gamma_{{\rm chg}}=1\ {\rm MHz}$
(representing the high-frequency limit of the noise spectra analyzed
in Ref. \citenum{Dial2013}) in Fig. \ref{fig:FminprimevsdeltaUab}(c)
{[}Fig. \ref{fig:FminprimevsdeltaUab}(d){]} and for $\Gamma_{{\rm chg}}=1\ {\rm GHz}$
in Fig. \ref{fig:FminprimevsdeltaUab}(e) {[}Fig. \ref{fig:FminprimevsdeltaUab}(f){]}.
We find that, while the gate fidelity for GaAs dots is further degraded
in the presence of fast charge noise, fidelities of up to $F_{\min}^{\prime\prime}\sim0.999$
are in principle possible for Si dots even in the presence of $1\ {\rm MHz}$
charge noise. Charge noise of frequency $\Gamma_{{\rm chg}}=1\ {\rm GHz}$
results in a significant reduction of the gate fidelity for Si dots,
which becomes similar to that for GaAs dots with $\Gamma_{{\rm chg}}=1\ {\rm GHz}$
{[}Fig. \ref{fig:FminprimevsdeltaUab}(e){]} as expected from the
fact that $\Gamma_{{\rm Si}}\ll1\ {\rm GHz}$ {[}see Fig. \ref{fig:phononSi}(a){]}. 

\section{Dependence of capacitive coupling on dot geometry}

\begin{figure}
\includegraphics[bb=20bp 50bp 320bp 620bp,width=2.2in]{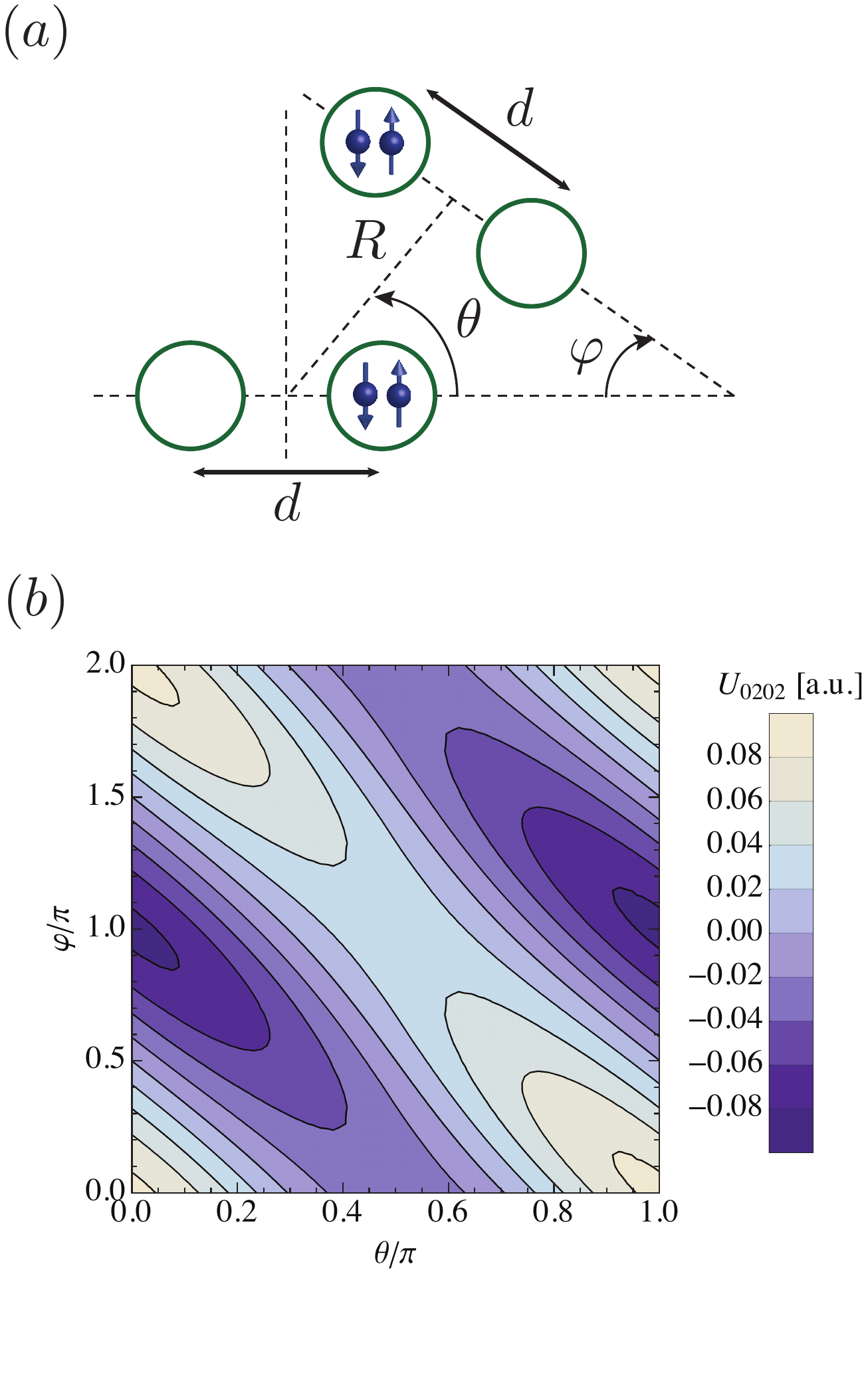}

\protect\caption{\label{fig:ccgeometry}(a) Illustration of a pair of double dots having
interdot distance $d,$ separated by distance $R,$ and with relative
in-plane orientation determined by the angles $\theta$ and $\varphi.$
(b) Variation of the capacitive coupling $U_{0202}$ {[}Eq. \eqref{eq:U0202}{]}
with $\theta$ and $\varphi$ for $R/d=3.$ }
\end{figure}

Finally, we consider how the capacitive coupling strength varies with
the relative orientation of the double dots \cite{Yang2011geometry}.
Specifically, we consider the geometry shown in Fig. \ref{fig:ccgeometry}(a)
and the general form of the capacitive interaction term in the Hamiltonian,
given by

\begin{equation}
H_{C}=\frac{1}{2}\sum_{i\neq j}U_{ij}n_{i}n_{j}.\label{eq:Hintfull}
\end{equation}
In Eq. \eqref{eq:Hintfull}, we have for notational convenience re-defined
the dot indices $a1,a2,b1,$ and $b2$ as $1,2,3,$ and $4,$ respectively.
The matrix element of the Coulomb interaction between the electrons
in dot $i$ and dot $j$ with center positions ${\bf R}_{i}=\left(x_{i},y_{i}\right)$
and ${\bf R}_{j}=\left(x_{j},y_{j}\right),$ respectively, is \cite{Mahan2000}
\begin{eqnarray}
U_{ij} & \equiv & \bra{ij}\frac{1}{r}\ket{ij}\nonumber \\
 & \equiv & \int\frac{\left|\Psi_{i}\left({\bf r}\right)\right|^{2}\left|\Psi_{j}\left({\bf r}^{\prime}\right)\right|^{2}}{\left|{\bf r}-{\bf r}^{\prime}\right|}d{\bf r}d{\bf r}^{\prime},\label{eq:Uijdef}
\end{eqnarray}
where $\Psi_{i}\left({\bf r}\right)\equiv\langle{\bf r}\ket{i}=\psi\left(x-x_{i}\right)\psi\left(y-y_{i}\right)$
and $\psi$ is the one-dimensional Gaussian function defined in Sec.
\ref{sec:phononrelaxation}. We assume $R\gg d$ {[}see Fig. \ref{fig:ccgeometry}(a){]}
and estimate $U_{ij}$ by the leading order term in the multipole
expansion of the Coulomb interaction as 
\begin{equation}
U_{ij}\sim\frac{1}{\left|{\bf R}_{i}-{\bf R}_{j}\right|}.\label{eq:Uijest}
\end{equation}
Defining $E\left(n_{1}\ n_{2}\ n_{3}\ n_{4}\right)$ as the energy
of the charge state $\ket{n_{1}\ n_{2}\ n_{3}\ n_{4}},$ the Coulomb
energy that sets the speed of the controlled-Z gate in the double
charge resonant regime is given by
\begin{eqnarray}
U_{0202} & \equiv & E\left(0202\right)-E\left(1111\right)-\left[E\left(0211\right)-E\left(1111\right)\right]\nonumber \\
 &  & -\left[E\left(1102\right)-E\left(1111\right)\right]\nonumber \\
 & = & E\left(0202\right)+E\left(1111\right)-E\left(0211\right)-E\left(1102\right)\nonumber \\
 & = & U_{13}+U_{24}-U_{14}-U_{23}.\label{eq:U0202}
\end{eqnarray}
We note that Eq. \eqref{eq:U0202} includes the term $U_{23}=U_{ab},$
which is the dominant term in $U_{0202}$ for the parameter regime
we consider in the present work. $U_{0202}$ depends on the parameters
$R,d,$ $\theta,$ and $\varphi$ through Eq. \eqref{eq:Uijest}. 

The dependence of $U_{0202}$ on the relative orientation of the two
double dots, determined by $\theta$ and $\varphi$ {[}see Fig. \ref{fig:ccgeometry}(a){]},
is shown in Fig. \ref{fig:ccgeometry}(b) for fixed $R/d.$ From this
dependence, we see that the linear geometry ($\theta=0,$ $\varphi=\pi$)
is associated with a minimum energy, corresponding to an attractive
dipole-dipole interaction of maximum strength, and therefore provides
the fastest gate. On the other hand, the case $\theta=0,$ $\varphi=0$
corresponds to a maximum repulsive interaction strength. 

Note that Eq. \ref{eq:Uijest} is approximately independent of the
dot size $\sigma.$ Thus, to leading order, the interqubit capacitive
coupling strength for $R\gg d$ is largely insensitive to variations
in the sizes of the dots and depends primarily on the dot center positions.
While the sensitivity of the intraqubit tunneling amplitudes to dot
size differences may modify the charge admixture $\eta_{\alpha}$
and thus lead to changes in the gate speed, this sensitivity will
not qualitatively affect the approach discussed in the present work.
In addition, knowledge of the dot size variation should in principle
enable tuning of the gate voltages controlling the double dot potentials
in order to compensate for changes in the intraqubit charge admixture
and thereby optimize the fidelity. 

\section{Conclusions}

In the present work, we have investigated capacitively coupled singlet-triplet
qubits in a pair of adjacent double quantum dots in the double charge
resonant regime, where the interqubit Coulomb interaction leads to
near-degeneracy between the $\ket{1111}$ and $\ket{0202}$ charge
states. This regime is different from that considered in Ref. \citenum{Taylor2005}
and subsequent work, where the two-qubit coupling relies on a repulsive
dipole-dipole interaction. Using the dynamics generated within the
two-singlet subspace by the capacitive coupling, we derived a sequence
for a controlled-phase gate that includes spin echo pulses to correct
for single-qubit dephasing. For this gate sequence, we showed that
rapid gates with fidelities greater than 0.9998 in the presence of
classical, static charge noise are in principle achievable by adjusting
the individual qubit detunings to appropriate values. We also studied
the relaxation of coupled singlet-triplet qubits via electron-phonon
interaction for quantum dots in both GaAs and Si. The full relaxation
rates, as well as their dipolar and quadrupolar contributions, were
calculated for both linear and purely quadrupolar dot geometries.
For the linear dot geometry, we showed that the presence of phonon-induced
decay results in a large decrease in the gate fidelity for GaAs dots
but does not significantly affect the fidelity in the case of Si dots
due to much slower charge relaxation. In addition, we found that fidelities
greater than 0.999 are in principle possible for Si dots even in the
presence of $1\ {\rm MHz}$ charge noise. Finally, we showed that
the linear geometry gives rise to the fastest two-qubit gate. 

These results demonstrate that the intraqubit detunings, interqubit
interaction strengths, and geometry of a capacitively coupled pair
of double dots can be chosen in order to optimize the controlled-Z
gate fidelity. Implementations of this gate in the double charge resonant
regime using Si dots arranged in a linear geometry should lead to
high fidelities in the presence of both quasistatic and fast charge
noise as well as relaxation via phonons. Improvements to the results
of the present work might be found by considering the double charge
resonant regime for, e.g., multi-electron singlet-triplet qubits \cite{Vorojtsov2004,Barnes2011,Higginbotham2014,Mehl2013},
which are expected to have enhanced robustness to charge noise due
to screening of the Coulomb interaction by the additional electrons
in the dots. Finally, we note that measured relations for the exchange
coupling as a function of detuning in double dots \cite{Maune2012,Dial2013}
deviate from the detuning dependence in Eqs. \eqref{eq:Ja} and \eqref{eq:Jb}
derived from the Hubbard model and thus may lead to different optimal
operating points for the controlled-phase gate. Potential future directions
therefore also include exploring extensions to the Hubbard model as
well as more sophisticated charge noise models \cite{Nielsen2012,Calderon-Vargas2015}
in order to obtain a more accurate description of capacitively coupled
double dots in the double charge resonant regime. 
\begin{acknowledgments}
We acknowledge useful discussions with A. Yacoby, S. Das Sarma, B.
Halperin, A. Pal, and S. Yang. We also thank M. Maghrebi and G. Solomon
for helpful comments. This work was supported by DARPA MTO and the
NSF-funded Physics Frontier Center at the JQI. 
\end{acknowledgments}

\appendix

\section{Basis states obtained via Schrieffer-Wolff transformation \label{sec:SW-basis-states}}

Here, we give expressions for the corrected states resulting from
the Schrieffer-Wolff transformation used to obtain the effective Hamiltonian
$H_{{\rm eff}}$ {[}Eq. \eqref{eq:Heff}{]}. Up to second order in
the charge admixture parameters $\eta_{a}$ and $\eta_{b}$ defined
in Sec. \ref{sec:Model}, we find
\begin{eqnarray}
\ket{\widetilde{S_{11},S_{11}}} & \approx & \left(1-\eta_{a}^{2}-\eta_{b}^{2}\right)\ket{S_{11},S_{11}}\nonumber \\
 & - & 2\eta_{a}\eta_{b}\frac{\left(U_{ab}^{2}-\delta^{2}+\Delta_{d}^{2}\right)}{\left(U_{ab}-\delta\right)^{2}-\Delta_{d}^{2}}\ket{S_{02},S_{02}}\nonumber \\
 & + & \sqrt{2}\eta_{b}\ket{S_{11},S_{02}}+\sqrt{2}\eta_{a}\ket{S_{02},S_{11}},\label{eq:S11S11}
\end{eqnarray}
\begin{eqnarray}
\ket{\widetilde{S_{02},S_{02}}} & \approx & \left[1-\eta_{a}^{2}\frac{\left(U_{ab}+\delta+\Delta_{d}\right)^{2}}{\left(U_{ab}-\delta-\Delta_{d}\right)^{2}}\right.\nonumber \\
 &  & \left.-\eta_{b}^{2}\frac{\left(U_{ab}+\delta-\Delta_{d}\right)^{2}}{\left(U_{ab}-\delta+\Delta_{d}\right)^{2}}\right]\ket{S_{02},S_{02}}\nonumber \\
 &  & -2\eta_{a}\eta_{b}\frac{\left(U_{ab}^{2}-\delta^{2}+\Delta_{d}^{2}\right)}{\left(U_{ab}-\delta\right)^{2}-\Delta_{d}^{2}}\ket{S_{11},S_{11}}\nonumber \\
 &  & +\sqrt{2}\eta_{a}\frac{\left(U_{ab}+\delta+\Delta_{d}\right)}{\left(U_{ab}-\delta-\Delta_{d}\right)}\ket{S_{11},S_{02}}\nonumber \\
 &  & +\sqrt{2}\eta_{b}\frac{\left(U_{ab}+\delta-\Delta_{d}\right)}{\left(U_{ab}-\delta+\Delta_{d}\right)}\ket{S_{02},S_{11}},\label{eq:S02S02}
\end{eqnarray}
\begin{eqnarray}
\ket{\widetilde{S_{11},S_{02}}} & \approx & \left[1-\eta_{a}^{2}\frac{\left(U_{ab}+\delta+\Delta_{d}\right)^{2}}{\left(U_{ab}-\delta-\Delta_{d}\right)^{2}}-\eta_{b}^{2}\right]\ket{S_{11},S_{02}}\nonumber \\
 &  & -2\eta_{a}\eta_{b}\frac{\left(U_{ab}^{2}+\delta^{2}-\Delta_{d}^{2}\right)}{\left(U_{ab}-\delta\right)^{2}-\Delta_{d}^{2}}\ket{S_{02},S_{11}}\nonumber \\
 &  & -\sqrt{2}\eta_{b}\ket{S_{11},S_{11}}\nonumber \\
 &  & -\sqrt{2}\eta_{a}\frac{\left(U_{ab}+\delta+\Delta_{d}\right)}{\left(U_{ab}-\delta-\Delta_{d}\right)}\ket{S_{02},S_{02}},\label{eq:S11S02}
\end{eqnarray}
\begin{eqnarray}
\ket{\widetilde{S_{02},S_{11}}} & \approx & \left[1-\eta_{a}^{2}-\eta_{b}^{2}\frac{\left(U_{ab}+\delta-\Delta_{d}\right)^{2}}{\left(U_{ab}-\delta+\Delta_{d}\right)^{2}}\right]\ket{S_{02},S_{11}}\nonumber \\
 &  & -2\eta_{a}\eta_{b}\frac{\left(U_{ab}^{2}+\delta^{2}-\Delta_{d}^{2}\right)}{\left(U_{ab}-\delta\right)^{2}-\Delta_{d}^{2}}\ket{S_{11},S_{02}}\nonumber \\
 &  & -\sqrt{2}\eta_{a}\ket{S_{11},S_{11}}\nonumber \\
 &  & -\sqrt{2}\eta_{b}\frac{\left(U_{ab}+\delta-\Delta_{d}\right)}{\left(U_{ab}-\delta+\Delta_{d}\right)}\ket{S_{02},S_{02}}.\label{eq:S02S11}
\end{eqnarray}

\section{\label{sec:Minimum-Fidelity}Minimum fidelity}

The expression in Eq. \eqref{eq:Fminintegrand} for the minimum fidelity
of the controlled $\pi$-phase (or controlled-Z) gate is obtained
using the particular state $\ket{\psi_{{\rm in}}}$ chosen for the
analysis in the present work. Here, we show that this initial state
represents only one possible element of a more general class of states
that minimize the gate fidelity (and thus maximize the error) for
a given charge noise distribution. We write $\hat{\rho}_{{\rm out}}^{\left(0\right)}\equiv U_{\phi}\ket{\psi}\bra{\psi}U_{\phi}^{\dagger}$
and $\hat{\rho}_{{\rm out}}\equiv U_{\phi}^{\prime}\ket{\psi}\bra{\psi}U_{\phi}^{\prime\dagger}$
for an arbitrary initial state $\ket{\psi}\equiv c_{TT}\ket{T_{11},T_{11}}+c_{ST}\ket{\tilde{S}_{11},T_{11}}+c_{TS}\ket{T_{11},\tilde{S}_{11}}+c_{SS}\ket{\widetilde{S_{11},S_{11}}}$,
where $c_{TT},c_{ST},c_{TS},$ and $c_{SS}$ are complex coefficients.
The gate fidelity then becomes {[}see Eq. \eqref{eq:Fmin}{]}
\begin{eqnarray}
f & \equiv & {\rm Tr}\left[\hat{\rho}_{{\rm out}}^{\left(0\right)}\hat{\rho}_{{\rm out}}\right]\nonumber \\
 & = & \left|\bra{\psi}U_{\phi}^{\dagger}U_{\phi}^{\prime}\ket{\psi}\right|^{2}\nonumber \\
 & = & A^{2}+B^{2}+2AB\cos\left[n\pi\left(\frac{j_{d}^{\prime}}{\Omega^{\prime}}-\frac{j_{d}}{\Omega}\right)\right],\label{eq:f}
\end{eqnarray}
with $A\equiv\left|c_{TT}\right|^{2}+\left|c_{SS}\right|^{2}$ and
$B\equiv\left|c_{TS}\right|^{2}+\left|c_{ST}\right|^{2}.$ Noting
that the minimum value of the cosine function is -1, we then find
$f{}_{\min}=A^{2}+B^{2}-2AB=\left(A-B\right)^{2}.$ This has a minimum
value of zero for $A=B.$ Together with the normalization condition
$A+B=1$ for $\ket{\psi},$ this yields $A=B=1/2,$ so that 
\begin{eqnarray}
f{}_{\min} & = & \frac{1}{2}+\frac{1}{2}\cos\left[n\pi\left(\frac{j_{d}^{\prime}}{\Omega^{\prime}}-\frac{j_{d}}{\Omega}\right)\right]\nonumber \\
 & = & \cos^{2}\left[\frac{n\pi}{2}\left(\frac{j_{d}^{\prime}}{\Omega^{\prime}}-\frac{j_{d}}{\Omega}\right)\right],\label{eq:fmingeneral}
\end{eqnarray}
which agrees with the expression for the minimum fidelity in Eq. \eqref{eq:Fminintegrand}
determined using the specific input state $\ket{\psi_{{\rm in}}}=\frac{1}{2}\left(\ket{T_{11},T_{11}}+\ket{\tilde{S}_{11},T_{11}}+\ket{T_{11},\tilde{S}_{11}}+\ket{\widetilde{S_{11},S_{11}}}\right).$
Note that for this state, $c_{TT}=c_{SS}=c_{TS}=c_{ST}=1/2,$ which
satisfies $A=B=1/2.$ Thus, $\ket{\psi_{{\rm in}}}$ represents a
particular initial state that minimizes the gate fidelity. All such
states lead to the same expression for $f_{\min}$ {[}Eq. \eqref{eq:fmingeneral}{]}.

\bibliographystyle{apsrev4-1}
\bibliography{DCRDQD}

\end{document}